\documentclass[]{interact}

\usepackage[dvips]{lscape}  

\usepackage[numbers,sort&compress,merge]{natbib}
\bibpunct[, ]{[}{]}{,}{n}{,}{,}

\usepackage{color}
\newcommand{\Blue}[1]{#1}

\begin{document}


\title{Spin-free orbital entropy, mutual information, and \Blue{correlation} analysis}

\author{
\name{Ji\v{r}\'{\i} Pittner\thanks{CONTACT J. Pittner. Email: jiri.pittner@jh-inst.cas.cz}}
\affil{J. Heyrovsk\'{y} Institute of Physical Chemistry, Academy of Sciences of the Czech \mbox{Republic, v.v.i.}, Dolej\v{s}kova 3, 18223 Prague 8, Czech Republic}
}

\maketitle

\begin{abstract}
Orbital entropies, pair entropies, and mutual information have become popular tools for analysis of strongly correlated wave functions.
They can quantitatively measure how strongly an orbital (e.g. from the DMRG active space) participates in the strong correlation 
and reveal the \Blue{correlation} pattern between different orbitals. 
However, this pattern can become rather complicated and sometimes difficult to interpret for large active spaces
and is not invariant with respect to the spin projection ($M_s$) component of the spin multiplet state. 
We introduce a modified spin-free orbital entropy, pair entropy, and mutual information, which  simplify the \Blue{correlation} analysis and are invariant with respect to $M_s$. 
By comparison of these quantities with their ``original'' spin-including counterparts
one can distinguish static correlation due to spin couplings from the ``genuine'' strong correlation  due to a multiconfigurational character of the wave function.
We illustrate the approach on a model consisting of a non-interacting dimer of triplet diradicals and
on a more realistic example of iron-sulfur bound complexes with one and two iron atoms.

\end{abstract}


\section{Introduction}

The problem of electron correlation has a central importance in quantum chemistry. 
The correlation effects\cite{sinanoglu1963} are often divided into two or three categories, although the division is somewhat fuzzy and not completely rigorous.
The dynamic (weak) correlation occurs due to pairwise Coulomb repulsion of the electrons when they come close to each other
and corresponds to many small contributions of excited Slater determinants to the wave function.
On the other hand, one speaks about strongly (or non-dynamically) correlated systems, when there is no single dominant Slater determinant, and two or more determinants have a comparable weight. 
Sometimes one distinguishes two sub-cases --- static correlation, where several Slater determinants with large coefficients are needed to
obtain correct spin symmetry, and ``true'' strong correlation which is related to quasidegenerate orbitals in molecules undergoing bond breaking, in transition metal complexes, etc.
For systems exhibiting dynamic (weak) correlation only, many single-reference methods exist, which are able to achieve a satisfactory ``chemical accuracy''.
For strongly correlated systems, the situation is much more difficult, since one has to accurately take into account both the strong and the dynamic correlation.
Traditionally, the presence of the strong correlation has been judged based on the magnitude of configuration interaction (CI) coefficients or the size of coupled cluster (CC) amplitudes. 
New tools for this purpose have been developed based on the notions of entanglement \Blue{and correlation}, which are central to quantum information theory (QIT)  \cite{Ziesche-1995,Nagy-1996,Nalewajski-2000,legeza2021fermionic,schilling2020H2,schilling2021jctc,schilling2024qst}.
The antanglement and correlation analysis employs the orbital entropy, which is a measure of \Blue{correlation} of a given orbital with all other orbitals, 
and the mutual information, which measures the \Blue{correlation} of two orbitals with each other, regardless of other orbitals.
\Blue{In this article we do not distinguish between quantum and classical correlation of the orbitals, which are both captured by the aforementioned quantities.}
The QIT-based approaches have become more popular in recent two decades \cite{Legeza-2003b,Legeza-2004a,Legeza-2004b,Legeza-2006a,Rissler-2006,Barcza-2011,Boguslawski-2012b,Boguslawski-2013,Barcza-2014,Boguslawski-2014,boguslawski2016prb,boguslawski2021jcp,boguslawski2022pccp},
since they are particularly suitable for use with the DMRG method, which also gained popularity in quantum chemistry, particularly
for studies of strongly correlated transition metal compounds \cite{Marti-2008a,Chan-2011,Chan-2012,Harris-2014,fe2s2chan2014,boguslawski2015cpl,gonzalez2015pccp,fe2s2veis2024}.
\Blue{
Among the important applications of the entropy-based measures belongs the optimization of orbital ordering, which is essential to
make the DMRG calculations efficient \cite{Legeza-2003b,Moritz-2005a}  and
the automated approach to active space selection  \cite{stein-reiher-2016}, which has the potential to turn the application of  multi-configurational methods into a black box approach.
Further development in this direction represents the Quantum Information-Assisted Complete Active Space Optimization \cite{schilling2023jpcl}, where besides the active space selection also an orbital optimization minimizing the out-of-CAS orbital correlation is performed.
}

Iron-sulfur clusters and complexes represent an important class of such compounds which is highly relevant in biochemistry \cite{fe2s2chan2014,fecomplexgeom,fe2s2complexgeom}.
They facilitate redox chemistry in living matter by carrying out electron transfer and catalytic  processes essential e.g. for nitrogen fixation, respiration, and photosynthesis. 
Their computational treatment represents a challenge due to the presence of strong correlation and many close lying states of different spin multiplicity
resulting from the partially occupied 3d-shells of iron atoms.
Although the high-spin state in these systems is dominated by a single configuration (but its lower-$M_s$ components are still multideterminantal),
the dynamic correlation can energetically favor the low-spin states, which typically have more complicated multiconfigurational character and represent thus a challenge to computational methods \cite{fe2s2veis2024,NeeseROHF2024,limanni2021jpca,dobrautz2021jctc}.
In this work we introduce spin-free orbital entropy and mutual information and illustrate their application to the \Blue{correlation} analysis in these systems.

\section{Theory}

Orbital \Blue{correlation} measures have been recognized as an important tool for analysis of correlated electronic wave functions
already in the nineties \cite{Ziesche-1995,Nagy-1996,Nalewajski-2000}
and have since become  widely used \cite{Legeza-2003b,Rissler-2006,Boguslawski-2012b,Boguslawski-2013,Boguslawski-2014},
hand-in-hand with the development of the DMRG method in quantum chemistry \cite{dmrg2023}.
\Blue{
They are, however, not restricted to any particular wave function ansatz and can be used with other strongly correlated methods as well, for example pCCD \cite{boguslawski2016prb,boguslawski2021jcp,boguslawski2022pccp}.
}
The central quantity in the quantum information theory is the von Neumann entropy defined as
\begin{equation}
\label{entropy}
S(\rho) = - {\rm Tr} (\rho \,\log \rho) = - \sum_p w_p\, \log w_p \;\;,
\end{equation}
where $\rho$ is the density matrix of the system (generally in a mixed state)
and $w_p$ its eigenvalues \cite{Nielsen-2000}.
In quantum information theory, usually one takes base-2 logarithm, while in quantum chemistry one usually employs natural logarithm.
For a system composed of subsystems A and B,
the reduced density matrix (RDM) of the subsystem A can be obtained by partially tracing the full density matrix over the ``environment'' B
\begin{equation}
\rho_A = {\rm Tr}_B \, \rho_{AB} 
\end{equation}
and used to compute the subsystem's entropy $S(\rho_A)$.
The RDM $\rho_B$ of the subsystem B  and its entropy can be obtained analogously.
Finally, the mutual information $I_{AB}$ between the two subsystems
can be defined as 
\begin{equation}
\label{mutualinformation}
I_{AB} = S(\rho_A) +S(\rho_B) -S(\rho_{AB})\;\;.
\end{equation}
Due to the sub-additivity of the  von Neumann entropy $I_{AB} \geq 0$ \cite{Nielsen-2000}. 
For pure states, the equality is achieved only if the 
state of AB is separable \Blue{(which for pure states means a product state)} and the mutual information can thus provide a measure of entanglement in this context.
However,  it can be nonzero for separable mixed states, reflecting their classical correlation.
\Blue{
In general, the mutual information mesures correlation of both classical and quantum nature.
Recently, additional quantum-information-theoretic measures have been introduced, which serve to distinguish the quantum and classical correlations \cite{williamson2010prl,schilling2021jctc,schilling2024faraday}.}

In quantum chemistry, these quantum information theory concepts are applied to molecular orbitals, which are considered 
as the ``subsystems''.
For this purpose, one can rewrite the most general form of molecular wave function
in a ``Fock-space'' expansion over direct product of all possible occupational states of $k$ orbitals
\begin{equation}
\label{wavefunction}
|\Psi\rangle = \sum_{n_1, n_2,\ldots,n_k=1}^4 C^{n_1,n_2,\ldots,n_k} |\psi_{n_1}\rangle \otimes|\psi_{n_2}\rangle \otimes  \ldots \otimes  |\psi_{n_k}\rangle
\;\;,
\end{equation}
where
the orbital states $|\psi_{n_i}\rangle$ span the 4-dimensional basis 
\begin{equation}
\label{orbitalbasis}
{\cal B} = \{|\rangle, |\uparrow\rangle, |\downarrow\rangle, |\uparrow\downarrow\rangle\}
\end{equation}
and $C^{n_1,\ldots,n_k}$ is the tensor of the expansion coefficients, which is typically rather sparse, combining only 
the terms of the same number of electrons and $S_z$ projection.
Note that unlike the usual expansion in Slater determinants, this wave function ansatz is not manifestly antisymmetric and the fermionic character is inherent in the expansion coefficient tensor.
Selecting an orbital number $i$ and considering the rest of the orbitals as an ``environment'' ${\cal E}_i= \{l: l=1,\ldots,k; l\neq i\}$ 
one can define the orbital RDM
\begin{equation}
\label{rho_i}
\rho_i = {\rm Tr}_{{\cal E}_i} |\Psi\rangle\langle\Psi|
\;\;,
\end{equation}
which is a $4\times4$ matrix and is diagonal due to the particle number and $S_z$ restrictions.
The orbital entropy can then be computed according to (\ref{entropy}), taking advantage of the diagonality of $\rho_i$
\begin{equation}
\label{orbitalentropy}
S_i \equiv S(\rho_i) = - {\rm Tr}_{\cal B} (\rho_i \ln \rho_i) =  - \sum_{p=1}^4 (\rho_i)_{pp}\, \ln (\rho_i)_{pp} \;\;
\;\;.
\end{equation}
The sum of all orbital entropies (in the active space of $k$ orbitals) is often called total \Blue{quantum information}
\begin{equation}
\label{totalentropy}
S_{\rm tot} = \sum_{i=1}^k S_i
\;\;.
\end{equation}
Similarly, for an orbital pair $i,j$ one can trace over the ``environment'' ${\cal E}_{ij}= \{l: l=1,\ldots,k; l\neq i \wedge l\neq j\}$
to define the pair RDM
\begin{equation}
\rho_{ij} = {\rm Tr}_{{\cal E}_{ij}} |\Psi\rangle\langle\Psi|
\;\;,
\end{equation}
which is a $16\times 16$ matrix, block-diagonal according to the particle numbers and $S_z$ values of the orbital pair states from ${\cal B}\otimes{\cal B}$.
The pair entropy is then analogously computed as
\begin{equation}
\label{pairentropy}
S_{ij} \equiv S(\rho_{ij}) = - {\rm Tr}_{{\cal B}\otimes{\cal B}} (\rho_{ij} \ln \rho_{ij}) =  - \sum_{p=1}^{16} \omega_p\, \ln \omega_p \;\;
\;\;,
\end{equation}
where $ \omega_p$ are the eigenvalues of $\rho_{ij}$ (we omit the indices $i,j$ at $\omega_p$ for simplicity).
\Blue{
The mutual information between orbitals $i$ and $j$ is then 
\begin{equation}
\label{pairmutual}
I_{ij} = S_i +S_j -S_{ij}
\;\;.
\end{equation}
Some works in quantum chemistry included in the definition of orbital mutual information (\ref{pairmutual}) a factor $1/2$ and a Kronecker $\delta$ term assuring that $I_{ii}=0$
\cite{Rissler-2006}, however, it is now customary to follow strictly the quantum information theoretic formula (\ref{mutualinformation}).
}
Using these quantities, a weighted complete graph on $k$ vertices can be constructed, $S_i$ being vertex weights and $I_{ij}$ serving
as edge weights.
A picture of this graph with weight coded by color and/or line thickness then illustrates the \Blue{correlation} in the molecule (for a particular choice of orbitals).
As a measure of the overall \Blue{correlation} in a given state, the sum of pairwise mutual informations can be employed
\begin{equation}
I_{\rm tot} = \sum_{i<j} I_{ij} 
\;\;,
\end{equation}
while for optimization of the orbital ordering in DMRG, it is useful to define the total correlation distance
\begin{equation}
I_{\rm dist} = \sum_{i<j} I_{ij}  (i-j)^2
\;\;.
\end{equation}

The orbital and pair RDMs can be expressed as expectations values using the 16 possible orbital basis transition operators $O^{ij}$  defined as
\begin{equation}
\psi_i = O^{ij} \psi_j\;\; \forall \psi_i , \psi_j \in {\cal B}
\;\;.
\end{equation}
The orbital RDM matrix elements are expectation values of these operators, while the pair RDM matrix elements can be expressed as expectation values of products of two such operators \cite{Rissler-2006,Boguslawski-2013}.
Since these operators can in turn be expressed as linear combinations of products of at most four fermionic creation or annihilation operators, 
the orbital and pair RDMs can be obtained for any wave function ansatz for which the corresponding 1-body to 4-body ``traditional'' reduced density matrices can be computed \cite{Boguslawski-2014}.
However, they are most conveniently obtained using the DMRG method which yields the wave function in the matrix product state (MPS) form \cite{Rommer-1997,Schollwock-2011},
which is an approximation of (\ref{wavefunction}) corresponding to the decomposition of the $C_{n_1,n_2,\ldots,n_k}$ tensor to a product of matrices with the bond dimension $M$
\begin{equation}
\label{mps}
 C^{n_1,n_2,\ldots,n_k}  = \sum_{a_1,a_2,\ldots,a_{k-1}=1}^M  A_{a_1}^{n_1}  A_{a_1,a_2}^{n_2} \cdots A_{a_{k-2}a_{k-1}}^{n_{k-1}} A_{a_{k-1}}^{n_{k}}
\;\;.
\end{equation}
The reduced density matrices can then be efficiently computed by contractions of the density matrix $|\Psi\rangle\langle\Psi|$ expressed using MPS \cite{Schollwock-2011}.
The \Blue{correlation} analysis thus became more popular in quantum chemistry hand in hand with the advent of DMRG.

Transition metal complexes, and particularly the multi-metallic ones, are examples of molecules with the most complicated strongly correlated
electronic structure. Even for a single electron configuration (a given occupation of the open shells) there are in general are many spin-adapted
configurations state functions (CSF) and even more Slater determinants. 
The \Blue{correlations} analysis as described above, is sensitive to the electron spin couplings, as the "up" and "down" singly occupied orbital
represent two different basis functions in ${\cal B}$.
This may lead to a rather complicated \Blue{correlation} diagram, which cannot distinguish which part of the correlation is ``static'' due to the spin coupling
and which is a ``genuine'' strong correlation.
In the following we present a modified orbital entropy, pair entropy and mutual information, which are spin-free, i.e. consider
a singly occupied orbital regardless of the direction of the spin of its electron.
We start with the orbital basis (\ref{orbitalbasis}) and consider the two singly occupied orbital states as a single indistinguishable ``microstate''
\begin{equation}
\label{orbitalbasis2}
\tilde{\cal B} = \{|\rangle, \{|\uparrow\rangle, |\downarrow\rangle\}, |\uparrow\downarrow\rangle\} \equiv  \{|0\rangle,  |1\rangle,  |2\rangle \}
\;\;,
\end{equation}
or, in another words, the orbital basis $\tilde{\cal B}$ only distinguishes the number of electrons in the orbital. 
The modified orbital  entropy is then computed as a trace of the spin-free orbital RDM $\tilde{\rho}_i$ over the spin-free orbital basis
\begin{equation}
\label{orbitalentropy2}
\tilde{S}_i = -{\rm Tr}_{\tilde{\cal B}} (\tilde{\rho_i} \ln \tilde{\rho_i}) = - \sum_{p=1}^3 (\tilde{\rho}_i)_{pp}  \ln (\tilde{\rho}_i)_{pp}
\;\;,
\end{equation}
where the nonzero diagonal elements of $\tilde{\rho}_i$ are
\begin{equation}
\label{rhotilde}
(\tilde{\rho}_i)_{11}= ({\rho}_i)_{11}, \;\; (\tilde{\rho}_i)_{22}= ({\rho}_i)_{22}+ ({\rho}_i)_{33},\;\; (\tilde{\rho}_i)_{33}= ({\rho}_i)_{44}
\;\;.
\end{equation}
Since the logarithm (with a base $>1$) is a monotonously increasing function, for $x,y>0$ it holds
 $(x+y)\ln(x+y) \geq x\ln x+y\ln y$ and thus $\tilde{S}_i \leq S_i$,
which is consistent with the fact that when computing $\tilde{S}_i$ we discard some information.
The equality would obviously occur in the atypical case if all electrons had the same spin (all $\alpha$ or all $\beta$).

The spin-free pair entropy is a bit more tricky - one cannot simply use the block-diagonal structure of $\rho_{ij}$, which happens to 
split into 9 blocks, which however, correspond to conservation of particle number and $S_z$. 
Taking the 5 ``super-blocks'' corresponding
to the particle numbers 0 to 4 is not consistent with the orbital entropy, the pair entropy would be underestimated and would not be equal 
to the sum of orbital entropies for two uncorrelated orbitals.
The basis, over which the pair entropy expression is traced, must be the direct product of the two spin-free orbital bases $\tilde{\cal B}$.
In analogy to (\ref{orbitalentropy2}), we thus define the spin-free pair entropy as
\begin{equation}
\label{pairentropy2}
\tilde{S}_{ij} = - \sum_{q=1}^9 \tilde{\omega}_q \ln \tilde{\omega}_q 
\;\;,
\end{equation}
where the summation runs over the nine spin-free basis states \Blue{$\tilde{\cal B}_q$} from $\tilde{\cal B} \otimes \tilde{\cal B}$, 
\Blue{i.e. $\tilde{\cal B}_q \in \{|00\rangle, |01\rangle,|02\rangle, |10\rangle,\ldots, |22\rangle \}$, where $|00\rangle$ abbreviates $|0\rangle\otimes |0\rangle$ etc.}. 
The quantities $\tilde{\omega}_q$ in (\ref{pairentropy2}) are the ``spin-summed eigenvalues''
 of $\rho_{ij}$
\begin{equation}
\label{spinsum2}
 \tilde{\omega}_q  = \sum_{p=1}^{16} \omega_p \sum_{k \in {\cal B}_q} c_{pk}^2
\;\;.
\end{equation}
Here  $\omega_p$ are the ``original'' eigenvalues of  $\rho_{ij}$ and $c_{pk}$ are the corresponding normalized eigenvectors, while 
$ {\cal B}_q$ denotes the subset of the pair basis $\cal B  \otimes \cal B$ which corresponds to the $q$-th element of  $\tilde{\cal B} \otimes \tilde{\cal B}$.
\Blue{ 
For example, for $\tilde{\cal B}_q=|1\rangle \otimes |2\rangle$, $ {\cal B}_q = \{|\uparrow\rangle\otimes |\uparrow\downarrow\rangle, |\downarrow\rangle\otimes \uparrow\downarrow\rangle \}$.
}

The quantity
$\tilde{\omega}_q$ thus represents a sum of  $\rho_{ij}$ eigenvalues weighted by the contribution of their eigenvector to a given spin-free basis state.
Clearly, $0 \leq  \tilde{\omega}_q \leq 1$ and
\begin{equation}
\sum_{q=1}^9  \tilde{\omega}_q  = \sum_{q=1}^9    \sum_{k \in {\cal B}_q}   \sum_{p=1}^{16} \omega_p  c_{pk}^2 = \sum_{k=1}^{16}  \sum_{p=1}^{16}  \omega_p  c_{pk}^2 = \sum_{p=1}^{16}  \omega_p  \sum_{k=1}^{16} c_{pk}^2 = 1
\;\;.
\end{equation}
The $\tilde{\omega}_q$ quantities can thus be interpreted as probabilities and employed in the entropy expression (\ref{pairentropy2}).
By the same argument as for orbital entropy, the spin-free pair entropy is not bigger than the original one $\tilde{S}_{ij} \leq S_{ij}$,
with equality if all electrons have the same spin.
The ``spin-summed eigenvalues'' (\ref{spinsum2}) might seem a bit ad-hoc construction, 
\Blue{
but actually it can be viewed as a generalization of the single-orbital spin-free entropy case (\ref{orbitalentropy2}), 
where the orbital RDM is diagonal, and hence the set of eigenvectors forms a unit matrix.
}
Moreover, when reconstructing $\rho_{ij}$ from its eigenvalues and eigenvectors as
\begin{equation}
(\rho_{ij})_{kl}   =  \sum_{p=1}^{16}   c_{pk}   \omega_p  c_{pl}
\end{equation}
 they can be alternatively viewed
as partial traces of the original pair RDM matrix over the ${\cal B}_q$ basis subset
\begin{equation}
\label{Omega_q}
 \tilde{\omega}_q  =  \sum_{p=1}^{16} \omega_p  \sum_{k \in {\cal B}_q}  c_{pk}^2  =  \sum_{k \in {\cal B}_q}   \sum_{p=1}^{16}   c_{pk}   \omega_p  c_{pk} = {\rm Tr}_{{\cal B}_q} (\rho_{ij})
\;\;.
\end{equation}
One might be tempted to introduce in this spirit a ``spin-summed pair RDM'' as
\begin{equation}
\label{spinsumwrong}
(\tilde{\rho}_{ij})_{rs} =   \sum_{k \in {\cal B}_r}  \sum_{l \in {\cal B}_s} (\rho_{ij})_{kl}
\end{equation}
and use its eigenvalues for the pair entropy definition. 
However, this spin summation does not conserve the trace ${\rm Tr} \tilde{\rho}_{ij} \neq {\rm Tr} \rho_{ij} =1$,
since some off-diagonal elements of $\rho_{ij}$ are added to the diagonal in the spin-free basis.
The eigenvalues of  $\tilde{\rho}_{ij}$  will thus in general not sum to one, so it is not a valid density matrix, 
corresponding to the fact that the spin summation (\ref{spinsumwrong}) is not a subsystem partial trace.

\Blue{
We should thus rather define the spin-free pair density matrix as being diagonal
\begin{equation}
\label{tilderho2}
(\tilde{\rho}_{ij})_{pq} =  \tilde{\omega}_q \delta_{pq} \;\;,
\end{equation}
where the indices $p,q$ run over the 9 elements of the basis  $\tilde{\cal B}_i \otimes \tilde{\cal B}_j$. 
The partial trace of this matrix over the spin-free basis of the orbital $j$ then recovers the spin-free orbital density matrix
\begin{equation}
\tilde{\rho}_i = {\rm Tr}_{\tilde{\cal B}_j} \tilde{\rho}_{ij} \;\;,
\end{equation}
since this trace combined with the trace in (\ref{Omega_q}) forms a complete trace over orbital $j$ combined with the spin summation over orbital $i$.
}
\Blue{Notice that other choices of the basis states for the spin-free two-orbital RDM, for example the 5 block space by total particle number and $M_s$, would not fulfill this requirement.}
When the two orbitals $i$ and $j$ are totally uncorrelated, $\rho_{ij} = \rho_i \otimes \rho_j$ and due to the diagonality of the orbital RDMs, $\rho_{ij}$  is also diagonal
with eigenvalues being products of the $\rho_i $ and $\rho_j$ diagonal elements, while the eigenvectors $c_{pk}$ form a unit matrix.
From (\ref{spinsum2}) then follows that the $ \tilde{\omega}_q$ quantities are products of the spin-summed orbital RDMs
\begin{equation}
\tilde{\omega}_{3p+q} = (\tilde{\rho}_i)_{pp}  (\tilde{\rho}_j)_{qq}
\end{equation} 
and the spin-free pair entropy then reduces to the sum of the spin-free orbital entropies
\begin{equation}
\tilde{S}_{ij} = -\sum_{p=1}^3 \sum_{q=1}^3 (\tilde{\rho}_i)_{pp}  (\tilde{\rho}_j)_{qq} [\ln (\tilde{\rho}_i)_{pp}  + \ln (\tilde{\rho}_j)_{qq}] = \tilde{S}_i + \tilde{S}_j
\;\;.
\end{equation}
It thus ``inherits'' the key properties of the original pair entropy and can be used to define the spin-free mutual information analogously to (\ref{pairmutual})
\Blue{
\begin{equation}
\label{pairmutual2}
\tilde{I}_{ij} = \tilde{S}_i +\tilde{S}_j -\tilde{S}_{ij}
\;\;.
\end{equation}
}
The spin-free mutual information is never bigger than the original one $\tilde{I}_{ij} \leq I_{ij}$. This is a consequence of the
data processing inequality from the classical information theory (cf. Theorem 2.8.1 in \cite{elements_information}).
It says that if $x,y,z$ are probability distributions and $z=f(y)$, then for the mutual information holds $I(x,y) \geq I(x,z)$.
Since the mutual information is symmetric, it can be applied twice and for $x'=f(x)$ and $y'=g(y)$ we have $I(x,y)  \geq I(x',y')$.
The orbital and  pair entropies and consequently the mutual information (\ref{pairmutual},\ref{pairmutual2}) are computed from the eigenvalues
of the density matrices, which fulfill all requirements on being interpreted as probability distributions, so the
classical information theory results hold for these quantities. Since the spin summation (\ref{spinsum2}) is a function mapping
between two such distributions, the data processing inequality must hold. Intuitively, it is clear that if we discard some
information by performing the spin summation, we cannot increase the amount of information correlating one orbital with another.
The transition from the original mutual information (\ref{pairmutual}) to the spin-free one (\ref{pairmutual2})
 can thus only simplify the diagrams for the \Blue{correlation} analysis.

\Blue{
Another advantage of the spin-free quantities is their invariance with respect to the spin component $M_s$ of a given spin multiplet state.
To show this, we assume that the same molecular orbitals have been employed in calculations of all the $M_s$ components of the multiplet
and start from the well known fact that the total $N$-electron wave function can be written as a product
of spatial and spin wave functions
\begin{equation}
\label{spinspatial}
\Psi = \Psi_{\rm P}(N,S) \Psi_{\rm S}(N,S,M_s) \;\;,
\end{equation}
which must belong to mutually adjoint irreducible representations of the permutation group ${\cal S}_N$ to maintain the total fermionic antisymmetry.
The spatial wave function $\Psi_{\rm P}(N,S)$ depends on the spin $S$, but is not dependent on the $M_s$ quantum number.
Since the spatial and spin coordinates represent independent integration variables, the trace is separable to spatial and spin degrees of freedom
\begin{equation}
\label{trace_separation}
{\rm Tr} |\Psi\rangle\langle \Psi| = ({\rm Tr_{\rm P}} | \Psi_{\rm P}(N,S) \rangle\langle  \Psi_{\rm P}(N,S)|)  ({\rm Tr_{\rm S}} | \Psi_{\rm S}(N,S,M_s) \rangle\langle  \Psi_{\rm S}(N,S,M_s)|)
\end{equation}
and this holds also for partial traces. 
The spin eigenfunctions are normalized, so the corresponding full trace is unity
\begin{equation}
\label{fullspintrace}
{\rm Tr_{\rm S}} | \Psi_{\rm S} \rangle\langle  \Psi_{\rm S}| = \langle  \Psi_{\rm S} | \Psi_{\rm S} \rangle = 1\;\;.
\end{equation}
Now, applying this to the partial trace in the definition of the orbital RDM (\ref{rho_i}) we have to consider four cases corresponding to the $i$-th orbital basis states (\ref{orbitalbasis}):
When the orbital $i$ is empty, we get
\begin{equation}
\label{caseempty}
(\tilde{\rho}_i)_{11} =(\rho_i)_{11} = \langle|({\rm Tr_{{\rm P}{\cal E}_i}}  | \Psi_{\rm P}(N,S) \rangle\langle  \Psi_{\rm P}(N,S)|) |\rangle \;\;,
\end{equation}
where we used (\ref{fullspintrace}), since all electrons are outside orbital $i$ and the spin trace over the ``environment'' ${\cal E}_i$ is the full spin trace.
Similarly, when the orbital $i$ is doubly occupied, the electrons in it must couple to the singlet spin, and the electrons in the ``environment'' are responsible to achieve the given spin state (or the wave function vanishes if this is not possible).
The spin eigenfunction is thus a product
\begin{equation}
\Psi_{\rm S}(N,S,M_s)  = \Psi_{{\rm S},i}(2,0,0)  \Psi_{{\rm S},{\cal E}_i}(N-2,S,M_s) 
\end{equation}
and due to the normalization (\ref{fullspintrace}) of $ \Psi_{{\rm S},{\cal E}_i}(N-2,S,M_s)$ we have
\begin{equation}
\label{casedouble}
(\tilde{\rho}_i)_{33} =(\rho_i)_{44} = \langle\uparrow\downarrow|( {\rm Tr_{{\rm P}{\cal E}_i}}  | \Psi_{\rm P}(N,S) \rangle\langle  \Psi_{\rm P}(N,S)|  . |\Psi_{{\rm S},i}(2,0,0)\rangle\langle\Psi_{{\rm S},i}(2,0,0)| ) |\uparrow\downarrow \rangle \;\;.
\end{equation}
These two matrix elements are thus clearly $M_s$-invariant. 
The situation is different for the singly occupied case.
Here the spin eigenfunction can be split into two components according to the spin in orbital $i$
\begin{equation}
 \Psi_{\rm S}(N,S,M_s)  = \alpha(i) \Psi'_{{\cal E}_i}(N-1) + \beta(i) \Psi''_{{\cal E}_i}(N-1)
\;\;,
\end{equation}
where $ \Psi'_{{\cal E}_i}(N-1)$ and $\Psi''_{{\cal E}_i}(N-1)$ are not in general spin eigenfunctions,
the spin-trace over ${\cal E}_i$ does not form a full spin trace and the $(\rho_i)_{22}$ and $(\rho_i)_{33}$ matrix elements are not invariant individually.
However, when taking their sum  $(\rho_i)_{22}+(\rho_i)_{33}$, 
we sum over the spin variable of the electron in orbital $i$ as well,
forming a full trace over all the spin variables 
\begin{equation}
{\rm Tr}_{{\rm S},i} {\rm Tr}_{{\rm S},{\cal E}_i} |\Psi_{\rm S}(N,S,M_s) \rangle\langle \Psi_{\rm S}(N,S,M_s) | = {\rm Tr}_{\rm S} | \Psi_{\rm S} \rangle\langle \Psi_{\rm S}| = 1
\end{equation}
and
\begin{equation}
\label{casesingle}
(\tilde{\rho}_i)_{22} = \langle 1| ({\rm Tr_{{\rm P}{\cal E}_i}}  | \Psi_{\rm P}(N,S) \rangle\langle  \Psi_{\rm P}(N,S)|) |1\rangle \;\;,
\end{equation}
where  $|1\rangle \in {\tilde{\cal B}}$ indicates orbital occupied by a single electron regardless of its spin.
The equations (\ref{caseempty},\ref{casedouble},\ref{casesingle})  for the individual cases  of orbital $i$ occupation can be combined, expressing the spin-free
orbital RDM as a spatial trace only
\begin{equation}
\tilde{\rho}_i = {\rm Tr_{{\rm P}{\cal E}_i}}  | \Psi_{\rm P}(N,S) \rangle\langle  \Psi_{\rm P}(N,S)| \;\;,
\end{equation}
which manifestly expresses the invariance of this quantity (and consequently of the spin-free entropy $\tilde{S}_i$) with respect to $M_s$.
Actually, this equation might serve as an alternative definition of the spin-free orbital RDM, which would be particularly suitable
in the context of spin-adapted methods like GUGA-CI.
For the pair reduced density matrix, the $M_s$ invariance can be shown analogously, analyzing the 16 cases corresponding to the pair basis ${\cal B}\otimes{\cal B}$.
}

Notice, however, that neither the spin-free nor the spin-dependent orbital entropies and mutual informations are invariant with respect to orbital transformations,
so the \Blue{correlation} pattern depends on the form of the orbitals (e.g. localization) \cite{schilling2023qst,schilling2024faraday,angeli2024}
and a sensible (chemically intuitive) choice can be very important for its interpretation \cite{Brandejs2019}.
\Blue{
This non-invariance can actually be turned into an advantage, simplifying the  multireference character of the wave function by
a suitable orbital transformation.
Of course, the orbital correlation cannot be fully transformed away by the change of on-particle basis, there exists an orbital basis for which it has a minimum value called particle correlation \cite{schilling2024faraday}.
This idea dates back to L\"{o}wdin \cite{lowdin1}, who has shown that natural orbitals give rise to most compact CI expansion, but the relationship to quantum information theory was uncovered only recently.
Legeza et al. have shown that by performing suitable orbital transformations (called fermionic mode optimization in DMRG \cite{legeza2016prl}), it is possible to minimize the multireference character 
(i.e. the orbital \Blue{correlation}) in the wave function \cite{legeza2023jmc}.
The orbital transformations are not limited to DMRG, but can be exploited in other strongly correlated methods as well.
Li~Manni et al. \cite{limanni2023jctc,dobrautz2022prb,limanni2021pccp,limanni2020jctc}
introduced a ``Quantum Anamorphosis'' approach in the spin-adapted GUGA-FCIQMC context and were able to achieve a dramatic compression
of the multireference wave function by localization of the orbitals carrying the unpaired spins.
They applied it to studies of polynuclear transition metal clusters, including the iron-sulfur compounds
\cite{limanni2021jpca,dobrautz2021jctc}, 
and by MO localization and reordering were able to achieve remarkable sparsity of the FCIQMC Hamiltonian matrices
leading to improvements in computational efficiency as well as insight into the spin structure of the investigated states.
Along similar lines, Schilling et al. \cite{schilling2024jpcl} have employed orbital transformations minimizing the total orbital correlation to improve the ability of the tailored  CCSD method to capture the correlation energy.
}

\section{Computational details}

For the illustration of the spin-free \Blue{correlation} analysis we employed a model example of a non-interacting dimer of CH$_2$ diradicals
in triplet ground states coupled to different resulting multiplicities. 
We took arbitrarily an equilateral  triangular geometry  with bond lengths 1.121\AA\  and angle 152.7$^\circ$ and computed FCI for the monomer (triplet ground state) employing canonical RHF orbitals in the STO-3G basis and freezing the 1s orbitals on carbon atoms. 
The dimer was constructed from two monomers separated by 100 a.u. distance, and through different choices of initial guess we converged the FCI to various linear combinations of the degenerate ground states with different spin multiplicities.
\Blue{For these small model systems} we employed our in-house FCI \Blue{(traditional string-based algorithm)} and entropy analysis C++ code.
\Blue{The molecular orbitals were visualized using CHARMOL \cite{charmol}.}

\begin{figure}[h]
\caption{
\label{fecomplex}
Structure of the  [Fe(SCH$_3$)$_4$]$^-$ and [Fe$_2$S$_2$(SCH$_3$)$_4$]$^{2-}$ complexes
}
\begin{center}
\begin{tabular}{ccc}
\includegraphics[width=6cm]{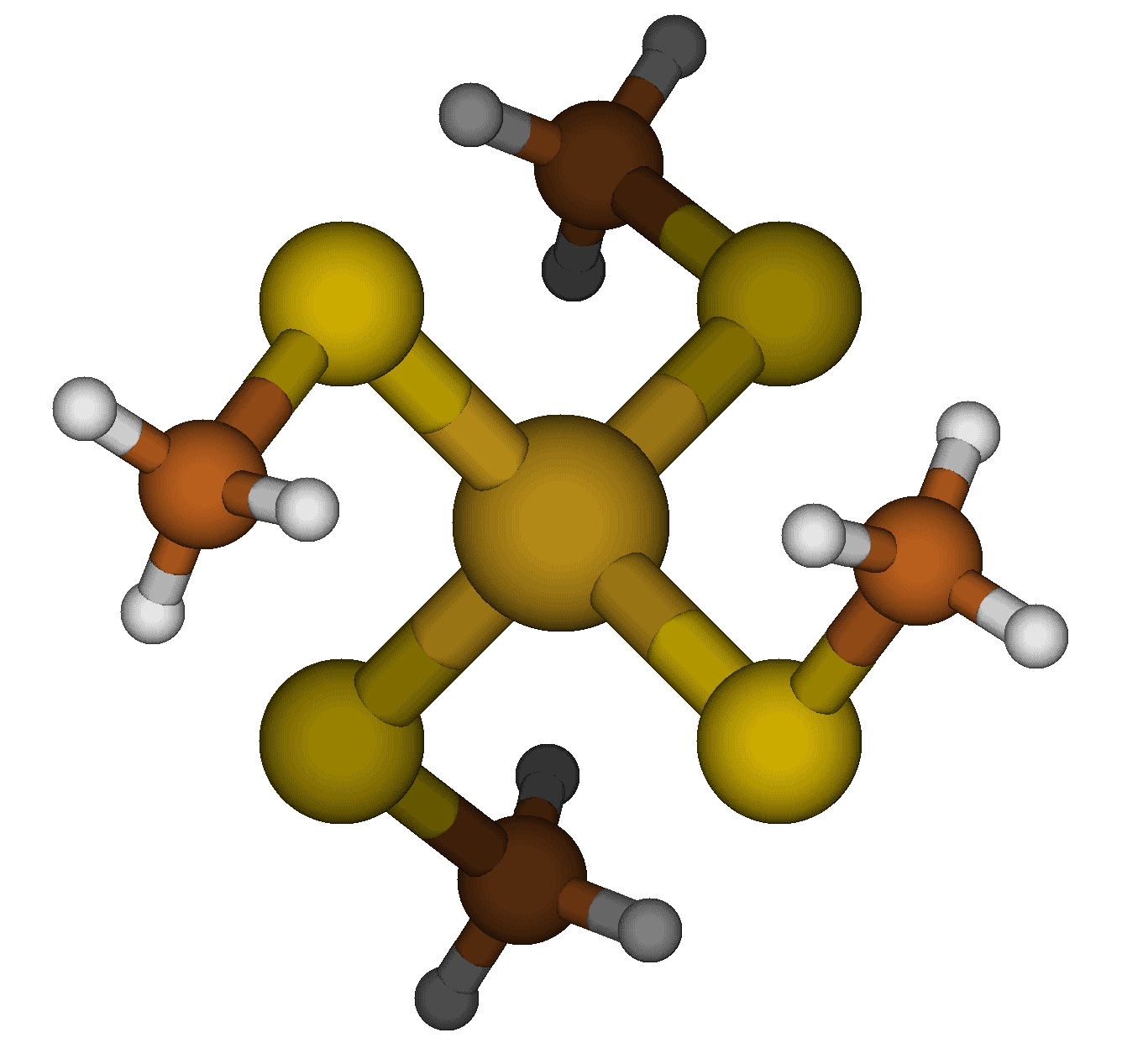} & $\;\;\;$ &\includegraphics[width=7cm]{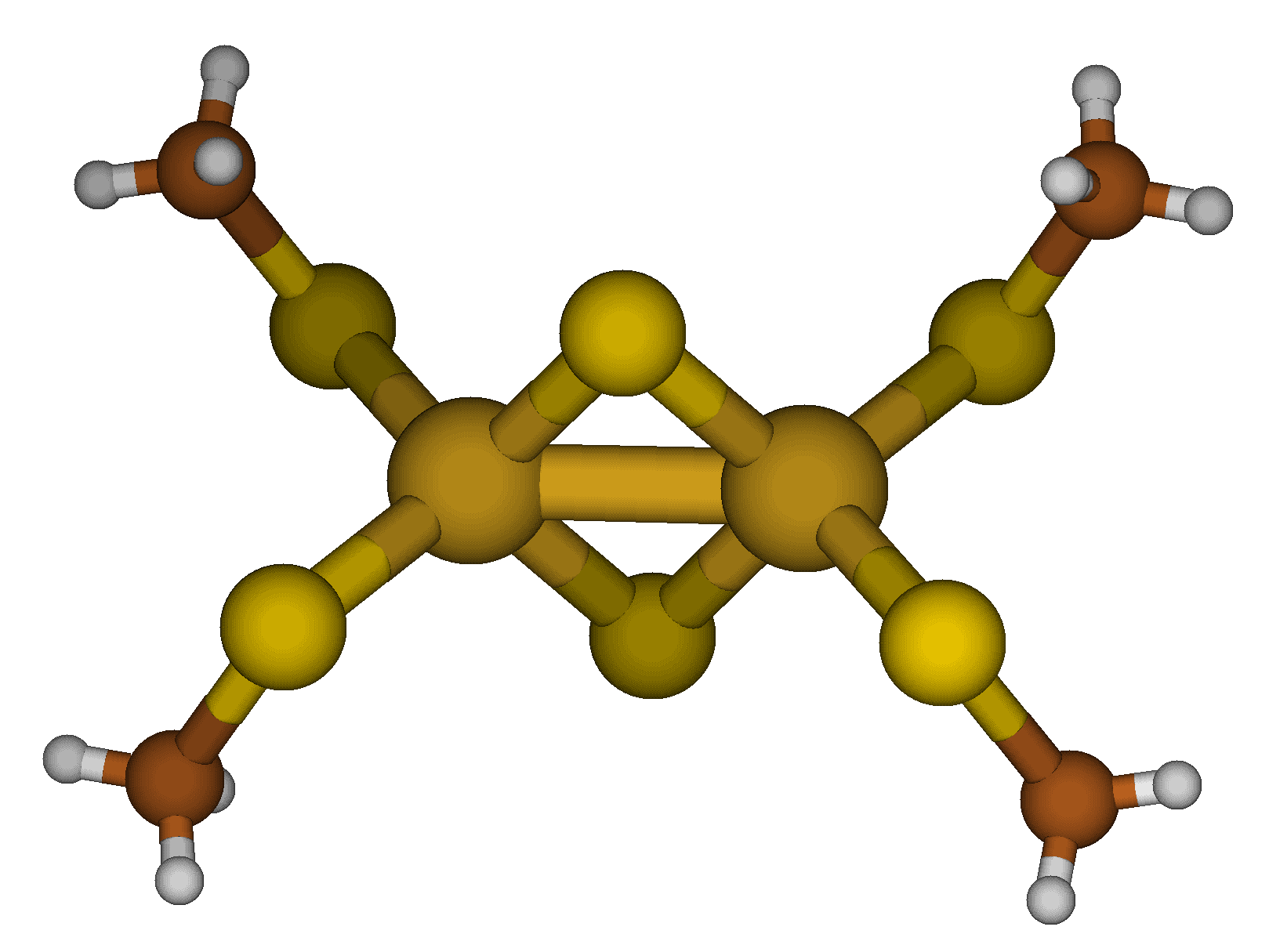}\\
\end{tabular}
\end{center}
\end{figure}

As a  more realistic application we have chosen the iron complexes [Fe(SCH$_3$)$_4$]$^-$ and [Fe$_2$S$_2$(SCH$_3$)$_4$]$^{2-}$ (cf. Fig.~\ref{fecomplex}), which were recently employed by Neese et al. to demonstrate the new restricted open shell SCF code in Orca \cite{NeeseROHF2024}.
The geometries of the complexes were originally reported in Refs.~\cite{fecomplexgeom,fe2s2complexgeom}, but are available in the Supplementary Information of  Ref.~\cite{NeeseROHF2024}.
We employed the Orca program \cite{Orca} and the Def2-TZVP basis set to perform a configuration-averaged ROHF SCF calculations and
integral dump, while the MOLMPS code\cite{molmps2020} was used for the subsequent DMRG calculations, and the new C++ code for the entropy analysis. 

For the single-iron complex, the ROHF SCF was done configuration-averaged for 5 electrons in the 5 d-orbitals, yielding one orbital set, 
\Blue{and separate ROHF calculations have been performed also for the three spin states.}
DMRG(13,14) calculations with bond dimension 512 were performed for states with multiplicities 2,4,6 \Blue{in each of the four orbital sets.} 
For the double-iron-sulfur complex, the configuration-averaged SCF was performed
for the 10 electrons in iron d-orbitals  and subsequently DMRG (14,14) and (22,21) calculations with bond dimension 2048 have been computed for the spin multiplicities 1,3,\ldots,11.
It turns out that in this complex the dynamic correlation is critical to get the correct energetic ordering of the states;
indeed, in the (14,14) the high-spin state is incorrectly the lowest, while already in the (22,21) space
the energy ordering was qualitatively correct (singlet ground state), in agreement with the post-DMRG and NEVPT2 calculations \cite{fe2s2veis2024}.
\Blue{We did not perform computations for several values of bond dimension and extrapolation to infinite bond dimension,
as we did not aim to compute the correlation energy with high accuracy, however, the active spaces and bond dimensions employed give a qualitatively correct energy ordering of the states and are sufficient for the qualitative correlation analysis.
The DMRG calculations were not spin adapted, but we checked the spin contamination  and the $S^2$ eigenvalues deviated from the correct ones
at most by 0.002 for all calculations of the single-iron complex and at most by 0.06 for the lowest root of each $M_s$ in the two-iron complex, and at most 0.2 for its higher roots (while the eigenvalues themselves are up to 30). }
However, it was not possible to converge all excited states corresponding to low $M_s$ components of the high $S$ states.

\section{Results}

\subsection{A non-interacting dimer of CH$_2$ diradicals}

As a simple illustrative example, we first investigated the  CH$_2$ diradical in its ground triplet state (at the chosen geometry) and its  non-interacting dimer.
The wave function of the monomer is dominated by a single Slater determinant in the $M_s=1$ component, or by a pair of Slater determinants
with equal weights in the $M_s=0$ component. For the non-interacting dimer, the spins can couple to energetically degenerate singlet, triplet, or quintet,
and a non-spin-adapted (determinantal-based) FCI code can converge to any linear combination thereof, which is compatible with the chosen $M_s$, depending on the initial guess. The wave function can thus vary between single-determinantal or multideterminantal character,
still describing the same physical system and yielding identical total energy.
The monomer's FCI energy was  -38.462451 a.u. and the dimer's energy was double that, to at least 6 decimals, in all calculations.

Table~\ref{ch2entropies} lists the orbital entropies computed from the FCI wave functions of the CH$_2$ diradical and its dimers,
for different values of the spin square operator $S^2$ and spin projection $M_s$. The spin-free entropies are given first,
with the ``traditional'' spin-including ones following in parentheses. 
Figure~\ref{ch2orbitals} shows the graphical representation of the orbitals.
The most interesting are the open shell (active) orbitals (3 and 4), which form the open shells of the monomer's triplet state.
It can be seen, that already for the monomer, the spin-free orbital entropies
are invariant to the choice of the  $M_s$ component of the triplet state, while the ``traditional'' entropies change their values.
In particular, the spin coupling of the active orbitals dramatically increases their entropy  with respect to the spin-free one.
Since we employed a determinant-based, not spin-adapted FCI code, for the dimer it did not always converge to an eigenfunction of $S^2$,
but to an arbitrary linear combination of the degenerate eigenfunctions. 
By using a random initial guess, we achieved a fractional value of $\langle S^2\rangle$, yet the spin-free entropies were not affected.
Moreover, for a spin-unpure state of the dimer the ``traditional'' entropies can differ for the two degenerate orbitals belonging to individual monomers, while the spin-free ones are the same.

\begin{landscape}
\begin{table}
\caption{\label{ch2entropies}Full-CI orbital entropies for the CH$_2$ diradical and its dimer. Spin-free entropies and original ones (in parenthesis) are listed. For the dimer, only one of the equal spin-free entropies of the equivalent monomer orbitals is listed, while the original entropies in parenthesis may have different values}
{
\begin{tabular}{|l|c|l|l|l|l|l|l|}
\hline
& & \multicolumn{6}{c|}{Orbital entropies}\\
\hline
System & $\langle S^2\rangle$, $M_s$ & DOCC 1 & DOCC 2 & ACTIVE 1 & ACTIVE 2 & VIRT 1 & VIRT 2\\
\hline
CH$_2$ &      2, 1   & 0.177 (0.194) & 0.115 (0.124) & 0.000 (0.040) & 0.181 (0.192) & 0.222 (0.238)& 0.092 (0.101)\\
CH$_2$ &      2, 0   & 0.177 (0.201) & 0.115 (0.124) & 0.000 (0.693) & 0.181 (0.849) & 0.222 (0.249)& 0.092 (0.101)\\
(CH$_2$)$_2$& 6, 2   & 0.177 (0.194) & 0.115 (0.124) & 0.000 (0.040) & 0.181 (0.192) & 0.222 (0.238)& 0.092 (0.101)\\
(CH$_2$)$_2$& 6, 1   & 0.177 (0.199) & 0.115 (0.124) & 0.000 (0.566) & 0.181 (0.724) & 0.222 (0.246)& 0.092 (0.101)\\ 
(CH$_2$)$_2$& 4, 1   & 0.177 (0.194,0.201) & 0.115 (0.124) & 0.000 (0.040,0.693) & 0.181 (0.192,0.849) & 0.222 (0.249,0.238)& 0.092 (0.101)\\
(CH$_2$)$_2$& 2, 0   & 0.177 (0.194) & 0.115 (0.124) & 0.000 (0.040) & 0.181 (0.192) & 0.222 (0.238)& 0.092 (0.101)\\
(CH$_2$)$_2$& 1.73, 0& 0.177 (0.196) & 0.115 (0.124) & 0.000 (0.255) & 0.181 (0.416) & 0.222 (0.241)& 0.092 (0.101)\\
\hline
\end{tabular}
}
\end{table}

\begin{figure}
\caption{\label{ch2orbitals}Valence molecular orbitals of the CH$_2$ diradical.}
\begin{tabular}{cccccc}
\includegraphics[width=3cm]{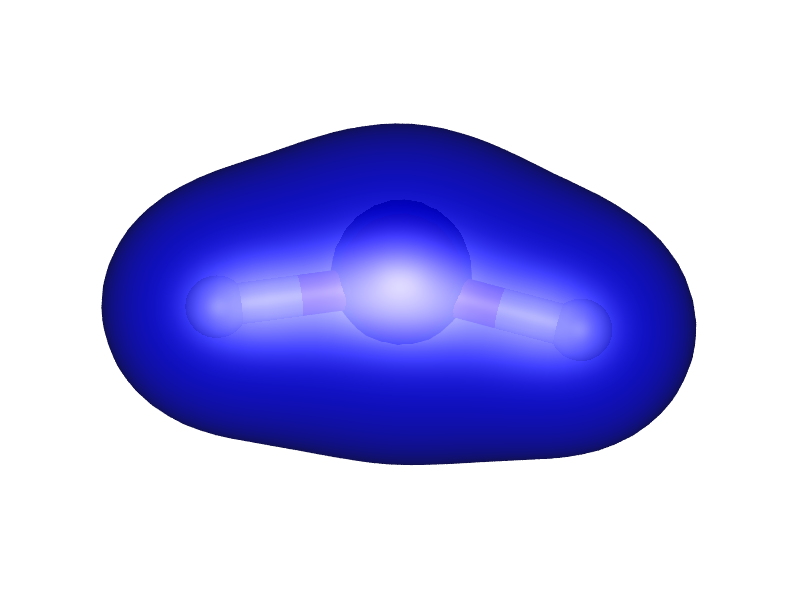} &
\includegraphics[width=3cm]{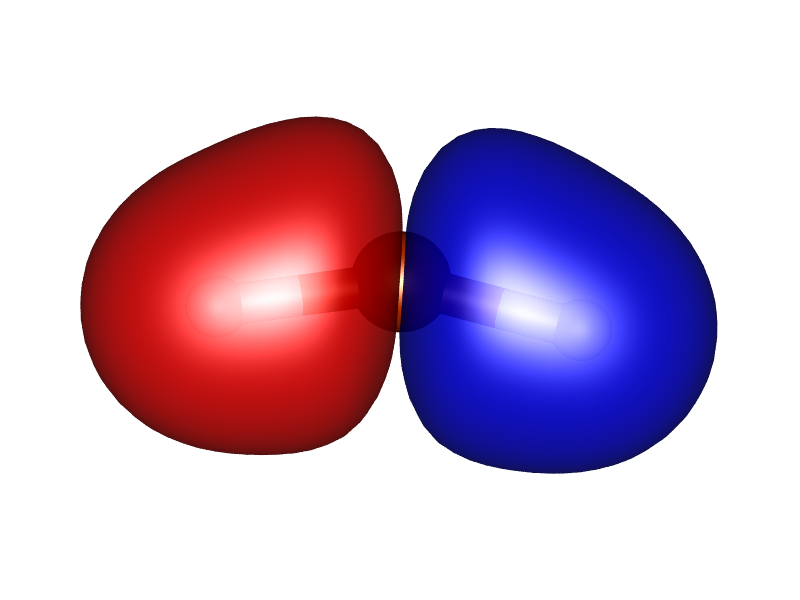} &
\includegraphics[width=3cm]{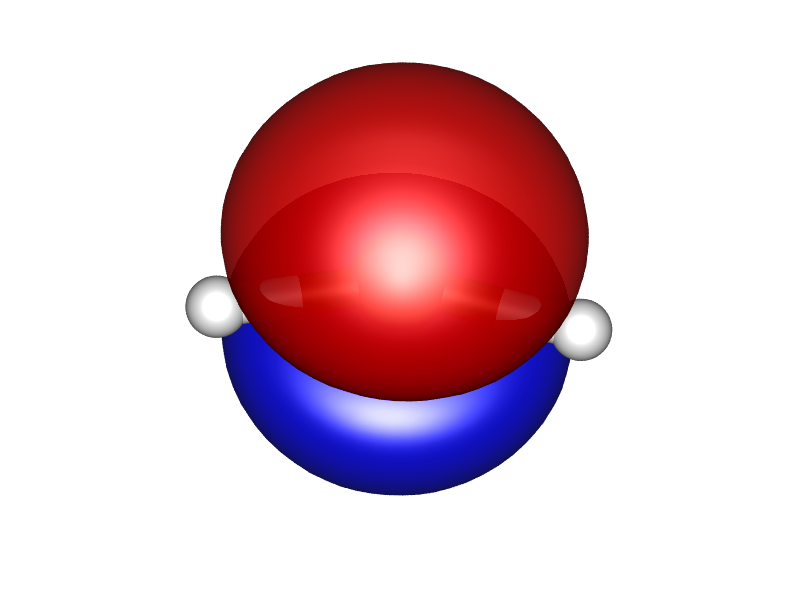} &
\includegraphics[width=3cm]{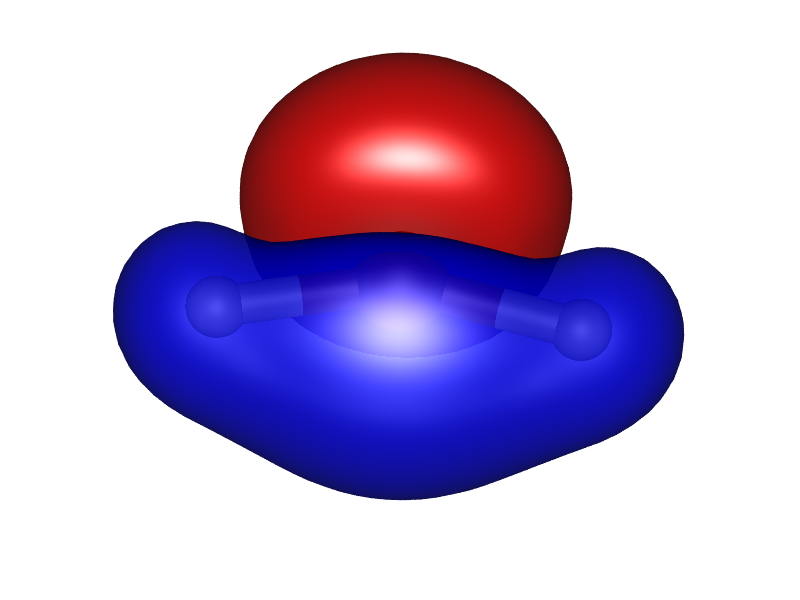} &
\includegraphics[width=3cm]{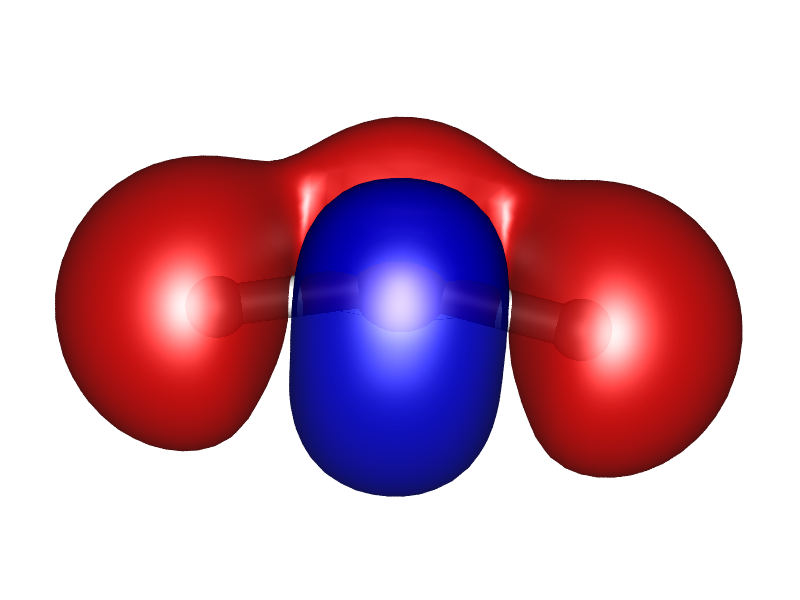} &
\includegraphics[width=3cm]{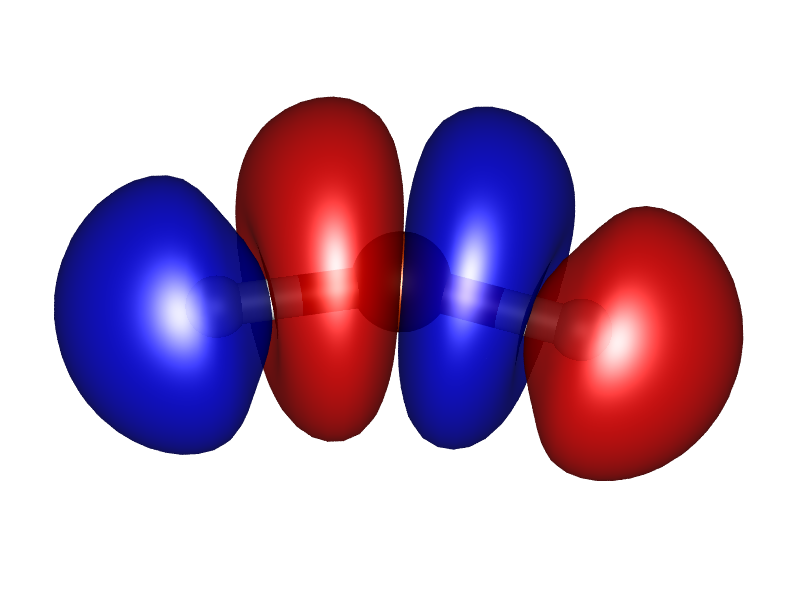} 
\\
 DOCC 1 & DOCC 2 & ACTIVE 1 & ACTIVE 2 & VIRT 1 & VIRT 2\\
\end{tabular}
\end{figure}
\end{landscape}

For the mutual information, there are too many data to show them all in a table, but it again turns out that the spin-free
mutual information is invariant to the  choice of spin state of the noninteracting dimer and is (up to a permutation of degenerate orbitals)
a direct sum of the mutual information matrices of the monomers, while the ``traditional'' mutual information does not have these properties.
The most interesting is the 4 x 4 submatrix corresponding to the two pairs of open shell orbitals of the monomers. The spin-free mutual information in this subspace vanishes, while the ``traditional'' one can acquire large values up to 0.31 for some spin couplings.
It is thus evident that the spin-free \Blue{correlation} analysis has some advantages, being useful at least as a supplementary to the ``traditional'' one.

\subsection{A complex with one iron atom}

As a first realistic example of the presented spin-free \Blue{correlation} analysis we turn to  the  [Fe(SCH$_3$)$_4$]$^-$ complex,
which contains an Fe(III) center in a S$_4$ symmetry coordination environment.
 The ground state follows the Hund's rule and turns to be a sextet, with each of the five MOs formed from 3d-orbitals of iron being singly occupied.
The DMRG calculations of the sextet for the three values of $M_s$ = 1/2, 3/2, and 5/2 yield energy -3011.13544(6-8) a.u., with differences in the order of micro Hartrees.
Reconstruction of the leading terms of the wave function from the MPS form confirmed that the $M_s=5/2$ component is single-determinantal,
while the $M_s=3/2$ component is dominated by 5 equally contributing Slater determinants, where four of the MOs corresponding to the iron 3d-orbitals have $\alpha$-spin and one has $\beta$ spin. For  $M_s=1/2$, ${5 \choose 2} = 10$ Slater determinants with three  $\alpha$ and two $\beta$ spins contribute equally.

\begin{figure}[h]
\caption{
\label{fecomplexanalysis}
\Blue{Correlation} analysis for the ground state ($S=5/2$) of the [Fe(SCH$_3$)$_4$]$^-$ complex, spin-including (left) and spin-free (right).
The bar graphs shows orbital entropies, while the color-coded size of the  mutual information matrix elements is displayed below.
The weighted graphs combining the orbital entropies and mutual information are plotted as well.
The top row contains results for  $M_s=5/2$ , middle row corresponds to  $M_s=3/2$  and the bottom one has  $M_s=1/2$.
In the correlation graph, the orbitals in their ascending energy order are placed in the clockwise direction starting from the gray indicated ``noon''.
\Blue{Configuration-averaged ROHF orbitals have been employed.}
}
\begin{center}
\begin{tabular}{ccc}
\includegraphics[width=8cm]{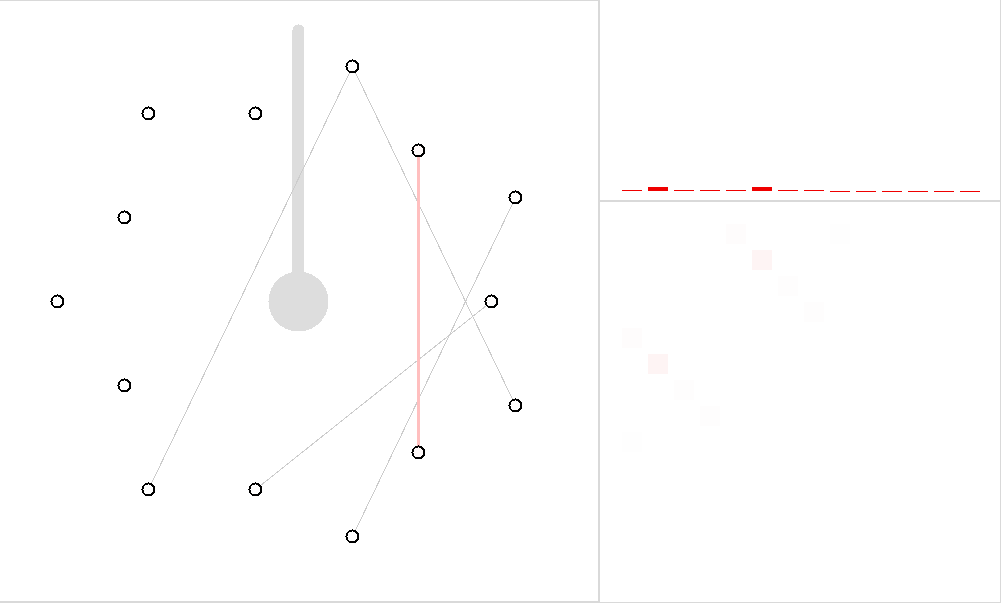} & $\;\;\;$ &\includegraphics[width=8cm]{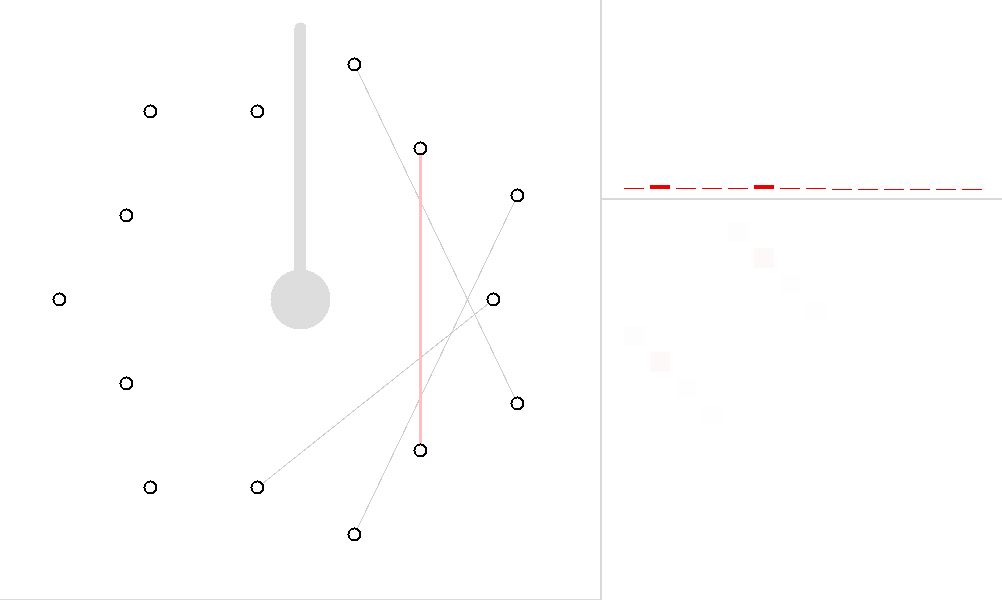}\\
\includegraphics[width=8cm]{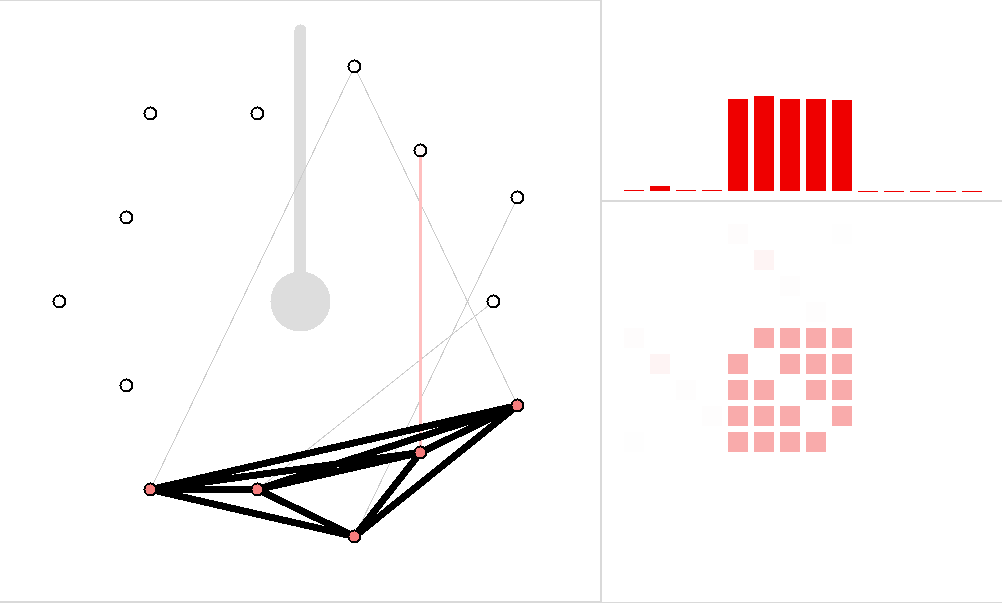} & $\;\;\;$ &\includegraphics[width=8cm]{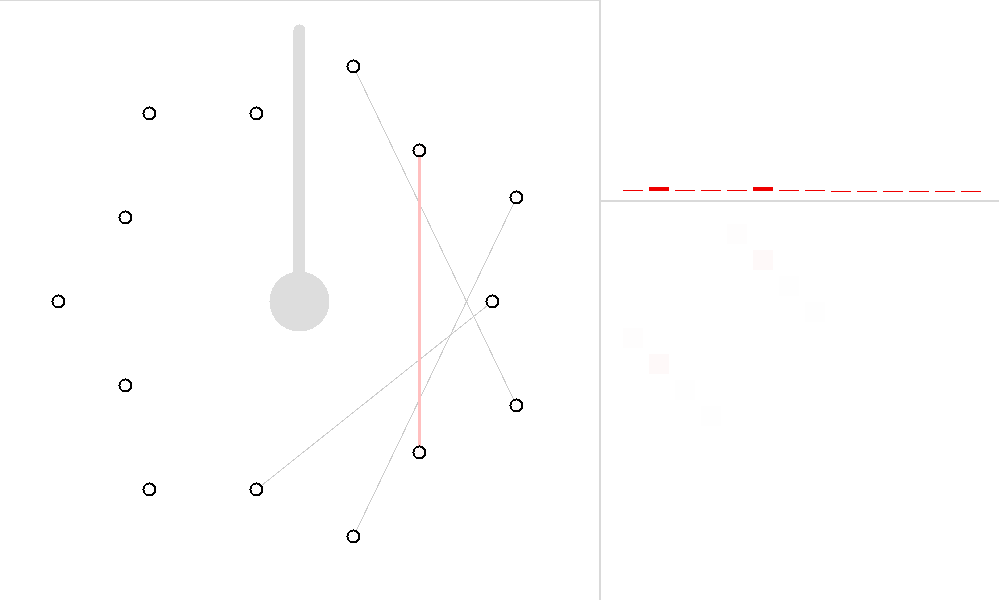}\\
\includegraphics[width=8cm]{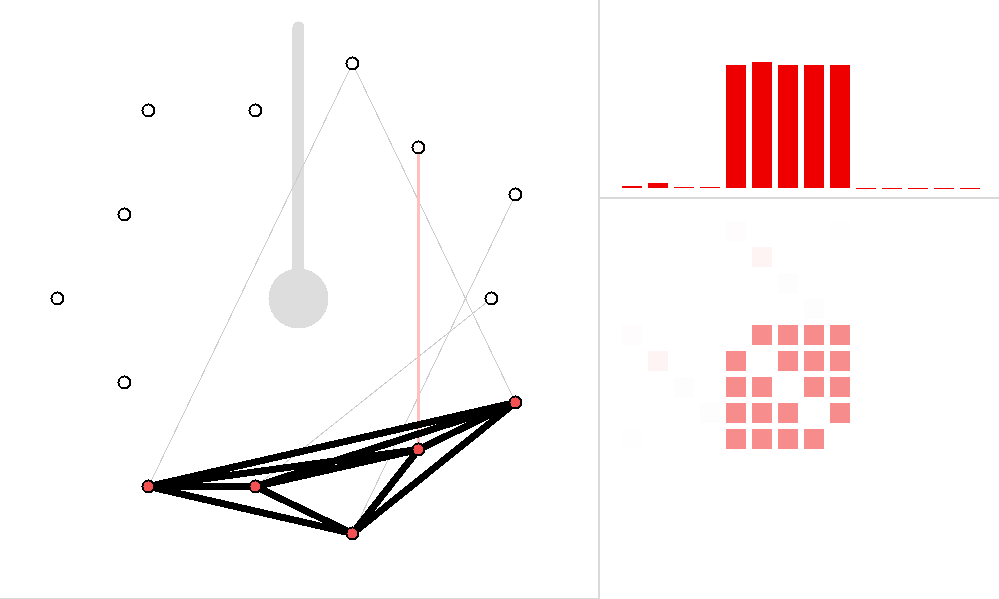} & $\;\;\;$ &\includegraphics[width=8cm]{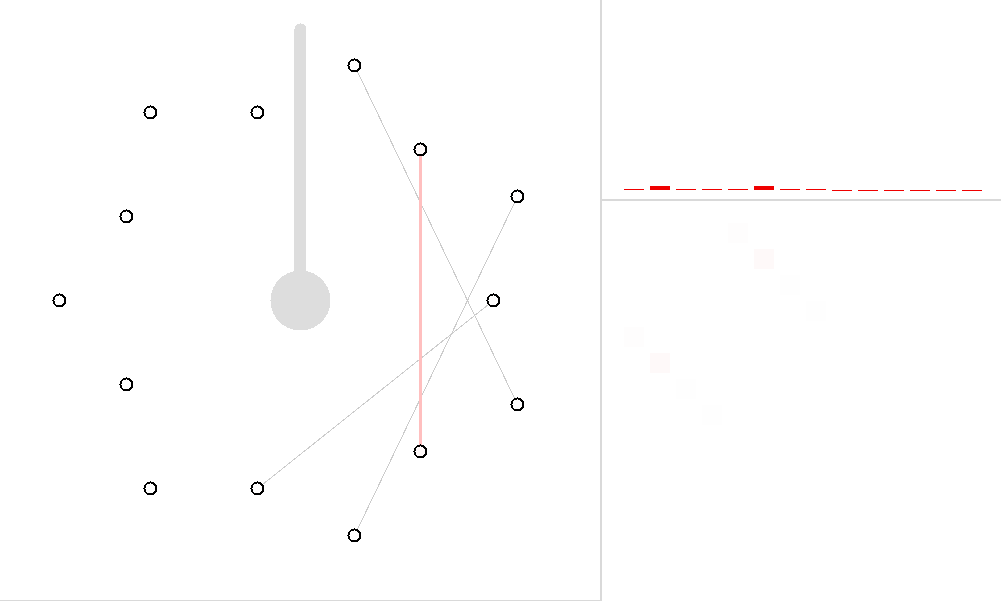}\\
\end{tabular}
\end{center}
\end{figure}

Figure~\ref{fecomplexanalysis} shows the \Blue{correlation} analysis for this state and its three positive $M_s$ components, in the spin-dependent and spin-free forms.
It can be seen that the three spin-free subfigures in the right column are identical, demonstrating the invariance of the spin-free
entropy and mutual information with respect to the $M_s$ value. The top left subfigure corresponds to the spin-dependent analysis 
for the  $M_s=5/2$  component, which is dominated by a single Slater determinant, and thus yield virtually identical results to the spin-free analysis.
On the other hand, the middle and bottom left subfigures for  $M_s=3/2$  and  $M_s=1/2$, respectively, show strong orbital entropies
for the MOs formed from 3d-orbitals of iron, and strong mutual information between each pair of them. For  $M_s=1/2$ the values are somewhat bigger,
in line with 10 Slater determinants contributing rather than five.

\begin{figure}[h]
\caption{
\label{root1fecomplexanalysis}
 \Blue{Correlation} analysis for the first excited state ($S=3/2$) of the [Fe(SCH$_3$)$_4$]$^-$ complex, spin-including (left) and spin-free (right).
The bar graphs shows orbital entropies, while the color-coded size of the  mutual information matrix elements is displayed below.
The weighted graphs combining the orbital entropies and mutual information are plotted as well. 
The top row contains results for $M_s=3/2$, while the bottom one has  $M_s=1/2$.
\Blue{Configuration averaged ROHF orbitals have been employed.}
}
\begin{center}
\begin{tabular}{ccc}
\includegraphics[width=8cm]{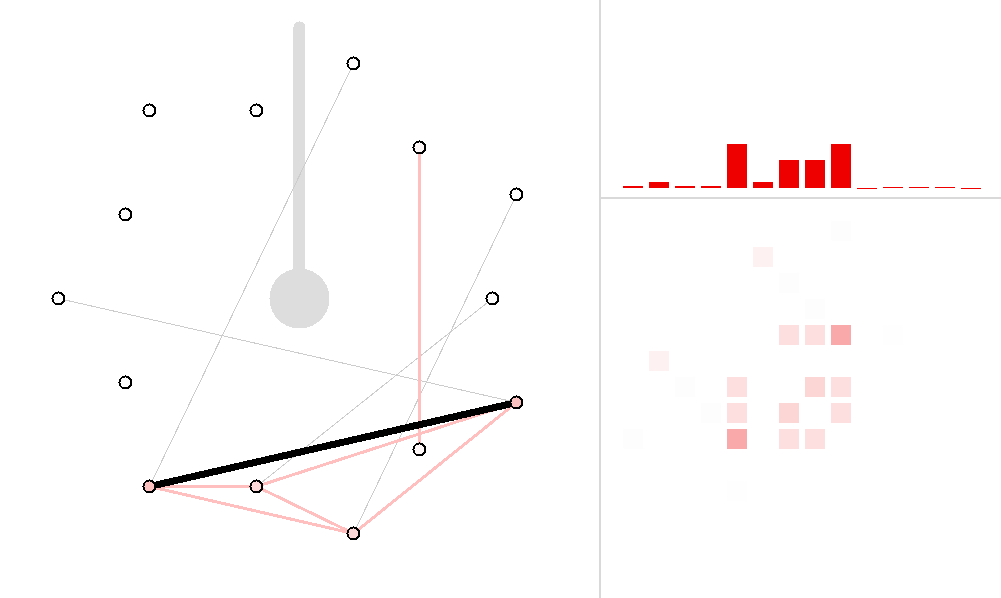} & $\;\;\;$ &\includegraphics[width=8cm]{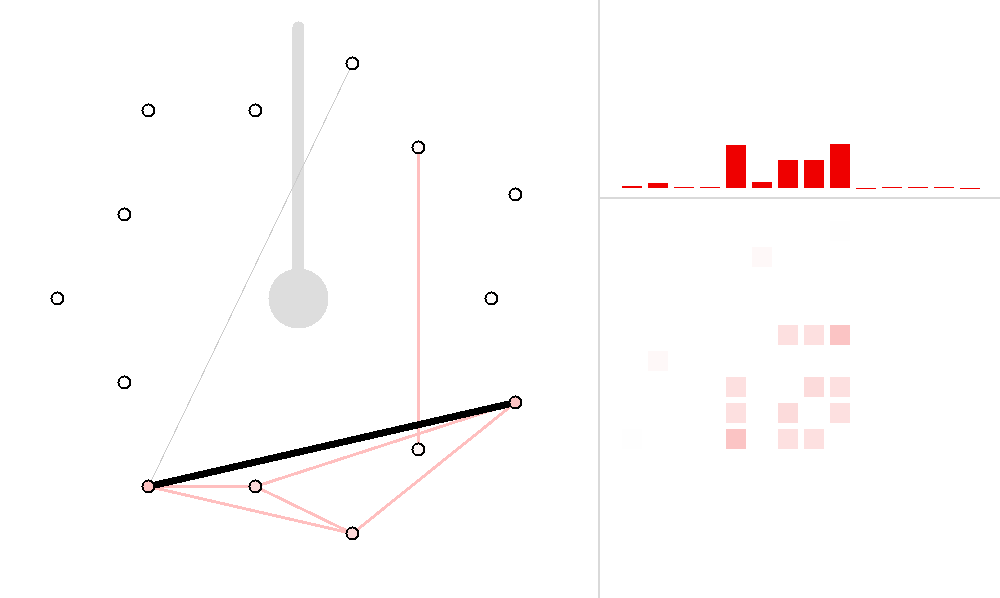}\\
\includegraphics[width=8cm]{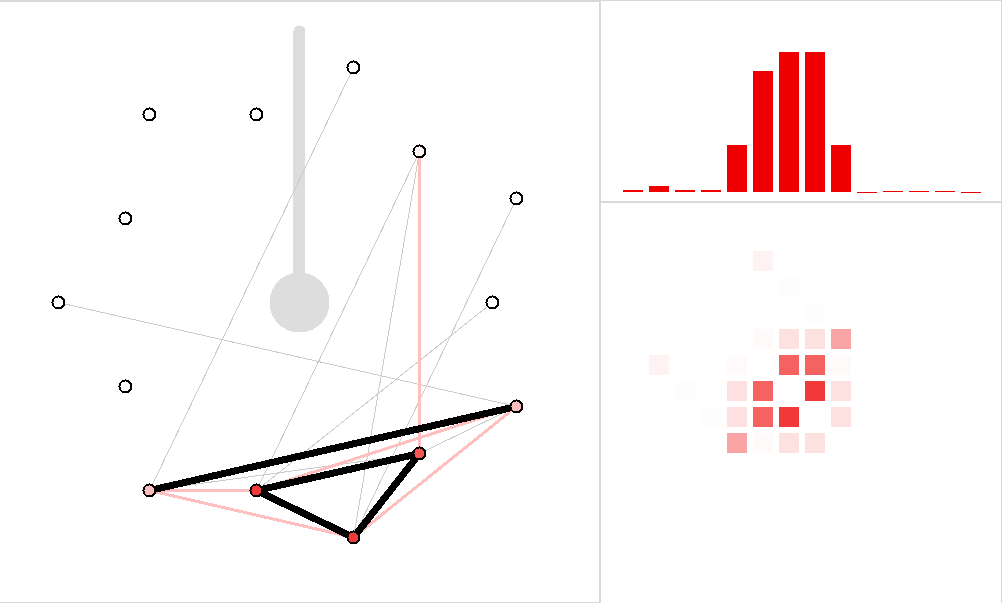} & $\;\;\;$ &\includegraphics[width=8cm]{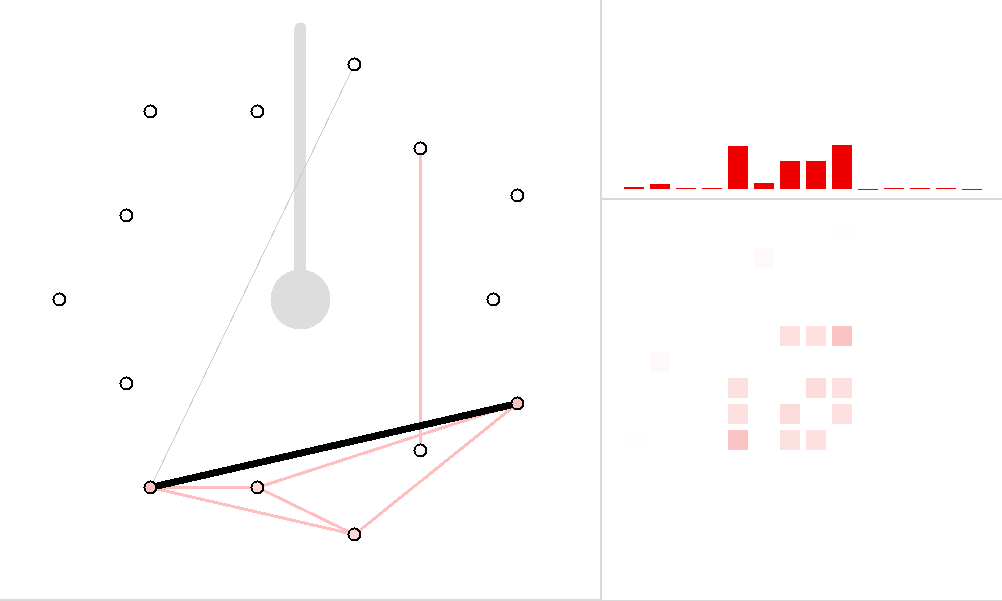}\\
\end{tabular}
\end{center}
\end{figure}

The approach can be used for excited states as well. For example, the first excited state with energy -3011.0279(49-50) a.u. is a quartet, with its  $M_s=3/2$  component dominated by the determinant $2\alpha\alpha\alpha0$ with a non-negligible contribution of  $0\alpha\alpha\alpha2$, while the  $M_s=1/2$ component has equal dominant contribution of determinants $2\beta\alpha\alpha0$, $2\alpha\beta\alpha0$, and $2\alpha\alpha\beta0$ with a non-negligible contribution of analogous $0\ldots2$ determinants.
Figure~\ref{root1fecomplexanalysis}  shows the corresponding   \Blue{correlation} analysis for the two positive  $M_s$ components of this state.
It can again be seen that the spin-free entropies and mutual informations are invariant to $M_s$, and they agree with the high-spin
spin-dependent case, showing strong entropies and mutual information for the first and last of the iron-3d dominated MOs.
The  \Blue{correlation} analysis for the $M_s=1/2$ component in the spin-dependent form shows in addition to this strong entropies and mutual information between the remaining three  iron-3d dominated MOs, which are purely a consequence of the spin coupling.

\begin{figure}[h]
\caption{
\label{root2fecomplexanalysis}
 \Blue{Correlation} analysis for the second excited state ($S=1/2$) of the [Fe(SCH$_3$)$_4$]$^-$ complex, spin-including (left) and spin-free (right).
The bar graphs shows orbital entropies, while the color-coded size of the  mutual information matrix elements is displayed below.
The weighted graphs combining the orbital entropies and mutual information are plotted as well.
\Blue{Configuration averaged ROHF orbitals have been employed.}
}
\begin{center}
\begin{tabular}{ccc}
\includegraphics[width=8cm]{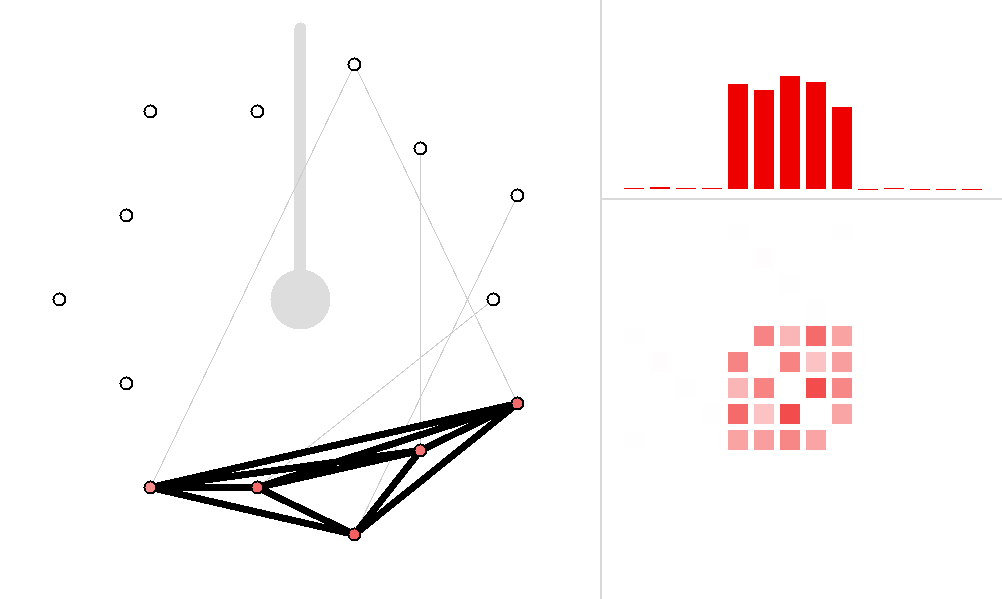} & $\;\;\;$ &\includegraphics[width=8cm]{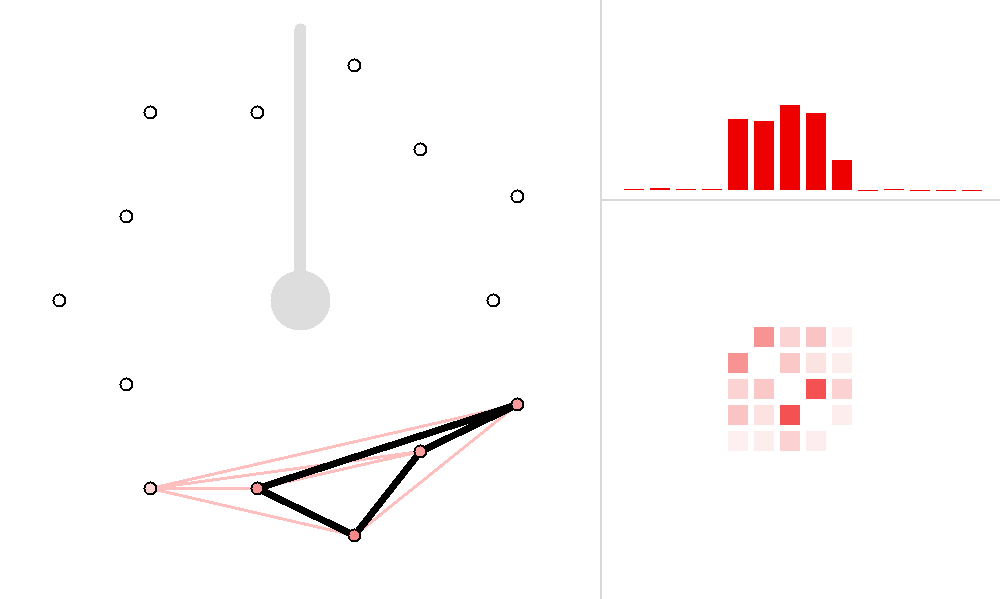}\\
\end{tabular}
\end{center}
\end{figure}

Figure~\ref{root2fecomplexanalysis}  shows the corresponding   \Blue{correlation} analysis for the second excited state, which is a doublet with energy -3011.016680 a.u..
Its wave function is dominated by several triples of equally contributing Slater determinants, where one of the 5 iron 3d MOs is doubly occupied, one empty, and the remaining three have two $\alpha$ and one $\beta$ electrons, in all three possible orderings. 
The static correlation due to spin couplings is thus inherently mixed with the strong non-dynamic correlation. 
This is reflected in the graphical analysis by the simplification of the mutual information graph when going from the spin-dependent to the spin-free analysis, and somewhat lower spin-free orbital entropies, particularly for one of the five orbitals.

\begin{table}
\caption{
\label{fecomplextable}
Spin-free and spin-including total \Blue{quantum informations} $\tilde{S}_{\rm tot}$, $S_{\rm tot}$ and mutual informations  $\tilde{I}_{\rm tot}$, $I_{\rm tot}$ for three states of the  [Fe(SCH$_3$)$_4$]$^-$ complex and their $M_s$ components, computed using configuration averaged ROHF orbitals.
}
\begin{tabular}{cccccc}
\hline
$S$ & $M_s$ &  $\tilde{S}_{\rm tot}$ & $S_{\rm tot}$ &  $\tilde{I}_{\rm tot}$ & $I_{\rm tot}$ \\
\hline
5/2 & 5/2 & 0.085 & 0.087 & 0.044 & 0.082 \\
5/2 & 3/2 & 0.085 & 2.588 & 0.044 & 3.354 \\
5/2 & 1/2 & 0.085 & 3.451 & 0.044 & 4.556 \\
3/2 & 3/2 & 0.879 & 0.891 & 0.898 & 1.082 \\
3/2 & 1/2 & 0.879 & 2.792 & 0.898 & 2.960 \\
1/2 & 1/2 & 3.709 & 5.623 & 2.208 & 4.372 \\
\hline
\end{tabular}
\end{table}

An overview of the spin-free and spin-including total \Blue{quantum informations} and mutual informations for the three states of this complex are given in Tab.~\ref{fecomplextable}.
As can be seen, the spin-free quantities are invariant with respect to $M_s$, while the spin-including ones strongly depend on the spin projection,
with a growing trend towards smaller $M_s$. 
This can be explained by the increased number of Slater determinants forming low-$M_s$ components of the spin eigenfunctions, resulting in increased static correlation.
However, even the spin-free total \Blue{quantum information} and mutual information have a growing trend with decreasing spin $S$, indicating more correlation
present in the lower-spin states.
This trend is somewhat hidden in the spin-including correlation measures due to their fluctuation with $M_s$, so in our view the
spin-free quantities  are a useful tool to uncover this dependence.

\begin{table}
\caption{
\label{fecomplextableotherorbs}
\Blue{
Spin-free and spin-including total quantum informations $\tilde{S}_{\rm tot}$, $S_{\rm tot}$ and mutual informations  $\tilde{I}_{\rm tot}$, $I_{\rm tot}$ for three states of the  [Fe(SCH$_3$)$_4$]$^-$ complex and their $M_s$ components, computed using ROHF orbitals optimized for different spin states.
}
}
\Blue{
\begin{tabular}{cc|cccc|cccc|cccc}
\hline
& &   \multicolumn{4}{|c|}{$S=5/2$ orbitals} &  \multicolumn{4}{c|}{$S=3/2$ orbitals} & \multicolumn{4}{c}{$S=1/2$ orbitals}\\
$S$ & $M_s$   &  $\tilde{S}_{\rm tot}$ & $S_{\rm tot}$ &  $\tilde{I}_{\rm tot}$ & $I_{\rm tot}$  &  $\tilde{S}_{\rm tot}$ & $S_{\rm tot}$ &  $\tilde{I}_{\rm tot}$ & $I_{\rm tot}$ &  $\tilde{S}_{\rm tot}$ & $S_{\rm tot}$ &  $\tilde{I}_{\rm tot}$ & $I_{\rm tot}$ \\
\hline
5/2 & 5/2 &    0.029 & 0.031 & 0.025 & 0.031 &    0.055 & 0.057 & 0.032 & 0.052 &    0.119 & 0.121 & 0.059 & 0.113 \\
5/2 & 3/2 &    0.029 & 2.532 & 0.025 & 3.303 &    0.055 & 2.557 & 0.032 & 3.325 &    0.119 & 2.622 & 0.059 & 3.385 \\
5/2 & 1/2 &    0.029 & 3.395 & 0.025 & 4.506 &    0.055 & 3.420 & 0.032 & 4.527 &    0.119 & 3.484 & 0.059 & 4.588 \\
3/2 & 3/2 &    0.892 & 0.911 & 0.859 & 1.066 &    0.854 & 0.866 & 0.881 & 1.054 &    0.913 & 0.925 & 0.919 & 1.118 \\
3/2 & 1/2 &    0.892 & 2.811 & 0.859 & 2.938 &    0.854 & 2.767 & 0.881 & 2.933 &    0.913 & 2.826 & 0.919 & 2.993 \\
1/2 & 1/2 &    3.662 & 5.584 & 3.281 & 5.694 &    3.617 & 5.532 & 2.487 & 5.163 &    3.740 & 5.653 & 2.938 & 5.390 \\
\hline
\end{tabular}
}
\end{table}

\Blue{
In order to investigate the influence of the underlying orbitals on the QIT measures we performed the DMRG calculations
of all the states in ROHF orbitals optimized for the individual spin states ($S=M_s= 5/2, \;3/2,\; 1/2$). The results are presented in Tab.~\ref{fecomplextableotherorbs}, which has the same structure as Tab.~\ref{fecomplextable}, repeated for the different orbitals.
As can be seen, the invariance of the spin-free measures with respect to $M_s$ holds for each set of calculations performed with the same orbitals, as expected.
Additionally, one can see that
the results for the sextet state are not much dependent on the orbitals, regardless of $M_s$, although the measures are lowest for its ``native''
ROHF orbital set.
For the quartet, the dependence on orbitals is also weak, although here some measures achieve minimal values in sextet orbitals.
The doublet behaves differently, its ``native'' ROHF orbitals do not minimize any of the QIT measures - for the total quantum information
the quartet orbitals are better, while for the total mutual information the configuration averaged orbitals perform best for this state.
A plausible explanation might be that the doublet state exhibits most ``genuine'' strong correlation (as can be seen from its spin-free QIT measures),
and the ROHF method, which includes the spin couplings but neglects this strong correlation, is thus not able to yield
``truly optimal'' orbitals for this state.
}

We have thus seen that for states of all spin multiplicities, the spin-free  \Blue{correlation} analysis was able to simplify the picture by filtering out the static correlation due to spin couplings.
In particular, the spin-dependent mutual information graph looks almost identical for the $M_s=1/2$ component of the doublet and sextet states, while their spin-free graphs are dramatically different, reflecting the different origin of the correlation.
It thus seems to be useful to accompany the ``traditional''  \Blue{correlation} analysis with its spin-free version.

\subsection{A complex with two iron atoms}

The  [Fe$_2$S$_2$(SCH$_3$)$_4$]$^{2-}$ complex is a  model of the biologically important iron-sulfur clusters, which act as
catalytic centers \cite{fe2s2complexgeom}. 
This species exhibits strong electronic correlation due to the coupling of the 3d-electrons
of the two iron atoms, while the dynamic correlation is very important to get the correct energetic order of the states of
different spin multiplicities (i.e. singlet ground state) \cite{fe2s2veis2024}.
In this work we did not aim to perform high-accuracy calculations of the energies of the states, rather we used the complex
as a non-trivial example for the  \Blue{correlation} analysis. 
Table~\ref{fe2s2energies} shows the energies of the $S=0-5$ states if this complex, computed in two different DMRG active spaces.
As can be seen, DMRG with a very small active space like e.g. (14,14)  yields inverted energy order of the states,
however,  DMRG(22,21) already yields a qualitatively correct ordering, although the excitation
energies are substantially underestimated with respect to DMRG-AC0 \cite{fe2s2veis2024} due to the lack of dynamic correlation. 

\begin{table}
\caption{
\label{fe2s2energies}
DMRG energies (in a.u.) of the [Fe$_2$S$_2$(SCH$_3$)$_4$]$^{2-}$ complex in different active spaces.
The energies of the $M_s=S$ components are given, the numerical errors in degeneracy were below $10^{-4}$ a.u.
and the errors in the $\langle S^2\rangle$ expectation values were below $10^{-3}$ except for $S=4$ with $M_s=1$, where it was 0.027.
}
{\small
\begin{tabular}{|l|c|c|c|c|c|c|}
\hline
DMRG space & $S=0$ & $S=1$ &$S=2$ &$S=3$ &$S=4$ &$S=5$ \\
\hline
(14,14)& not.conv. & -5068.002263 & -5068.003314 & -5068.004856 & -5068.0069483 & -5068.010014\\
(22,21)&-5068.033555 & -5068.033449 & -5068.033228 & -5068.032824 & -5068.032309 & -5068.032140\\
\hline
\end{tabular}
}
\end{table}

\begin{figure}[h]
\caption{
\label{S0analysis}
 \Blue{Correlation} analysis for the ground ($S=0$) state of the [Fe$_2$S$_2$(SCH$_3$)$_4$]$^{2-}$ complex, spin-including (left) and spin-free (right).
The bar graphs shows orbital entropies, while the color-coded size of the  mutual information matrix elements is displayed below.
The weighted graphs combining the orbital entropies and mutual information are plotted as well.                                         }
\begin{center}
\begin{tabular}{ccc}
\includegraphics[width=8cm]{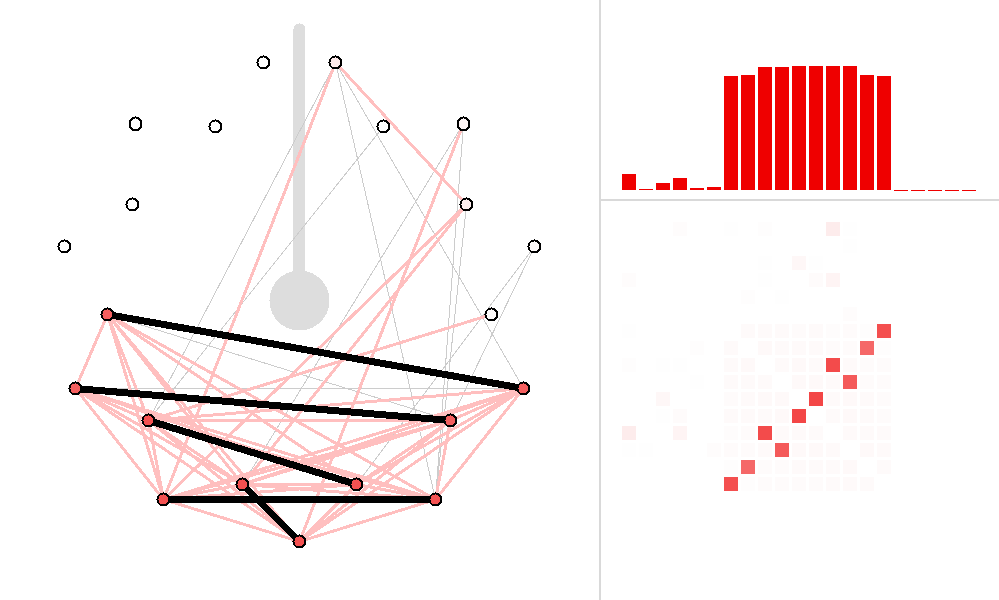} & $\;\;\;$ &\includegraphics[width=8cm]{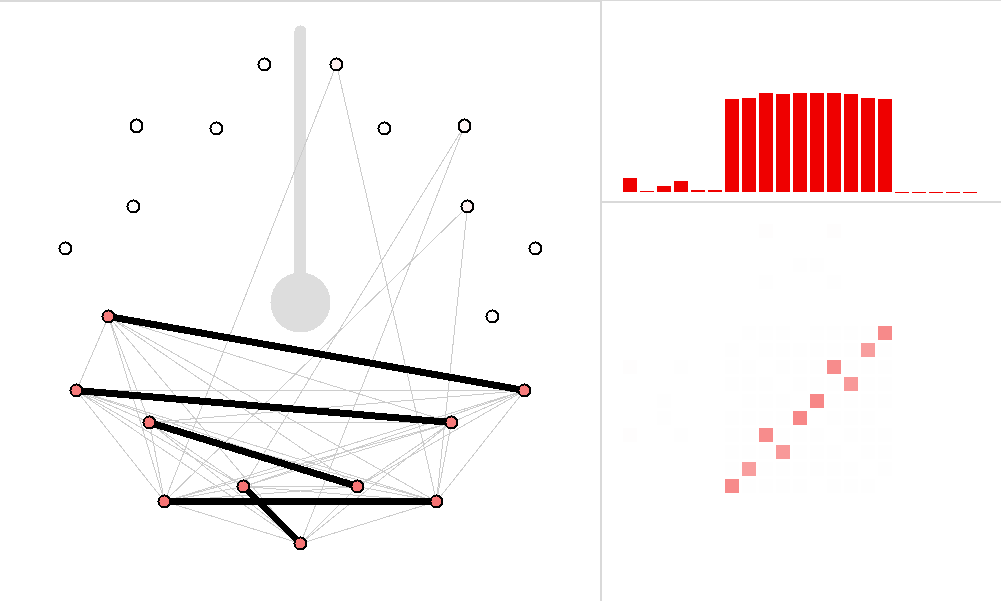}\\
\end{tabular}
\end{center}
\end{figure}

Fig.~\ref{S0analysis} shows the  \Blue{correlation} analysis for the ground ($S=0$) state. It clearly shows the
presence of the full pairwise ``anti-ferromagnetic'' coupling between the pairs formed from the 3d-dominated MOs of the two iron atoms.
Smaller contributions in the wave function come  from determinants with some open shells, which have to spin-couple to singlets,
and this is reflected by the weak mutual information lines, which disappear in the spin-free subfigure on the right, and the slightly lower spin-free orbital entropies.

\begin{figure}[h]
\caption{
\label{S1analysis}
 \Blue{Correlation} analysis for the first excited ($S=1$) state of the [Fe$_2$S$_2$(SCH$_3$)$_4$]$^{2-}$ complex, spin-including (left) and spin-free (right).
The bar graphs shows orbital entropies, while the color-coded size of the  mutual information matrix elements is displayed below.
The weighted graphs combining the orbital entropies and mutual information are plotted as well.
The $M_s=1$ component is in the top row, with $M_s=0$ component below.
}
\begin{center}
\begin{tabular}{ccc}
\includegraphics[width=8cm]{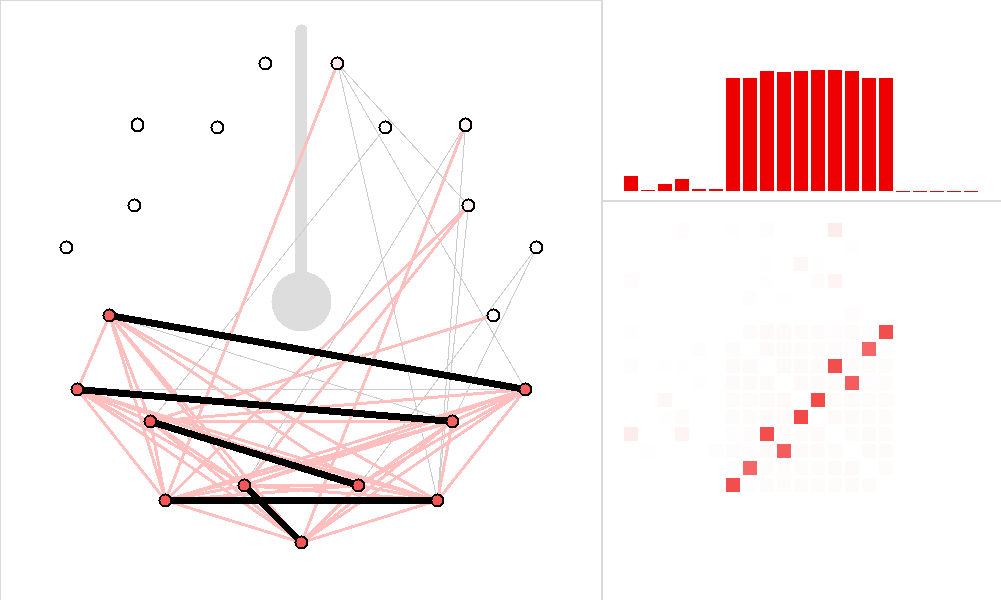} & $\;\;\;$ &\includegraphics[width=8cm]{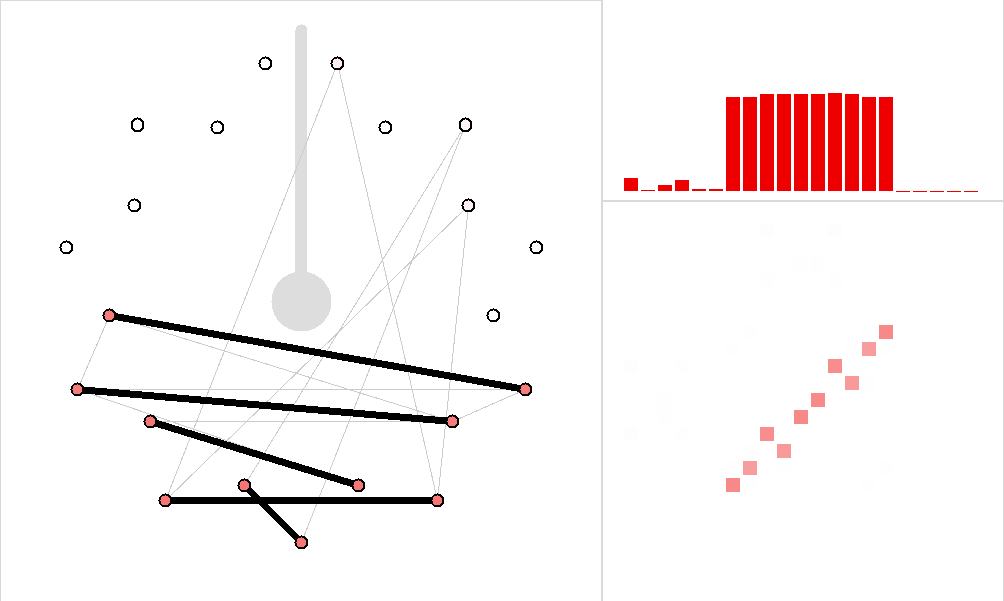}\\
\includegraphics[width=8cm]{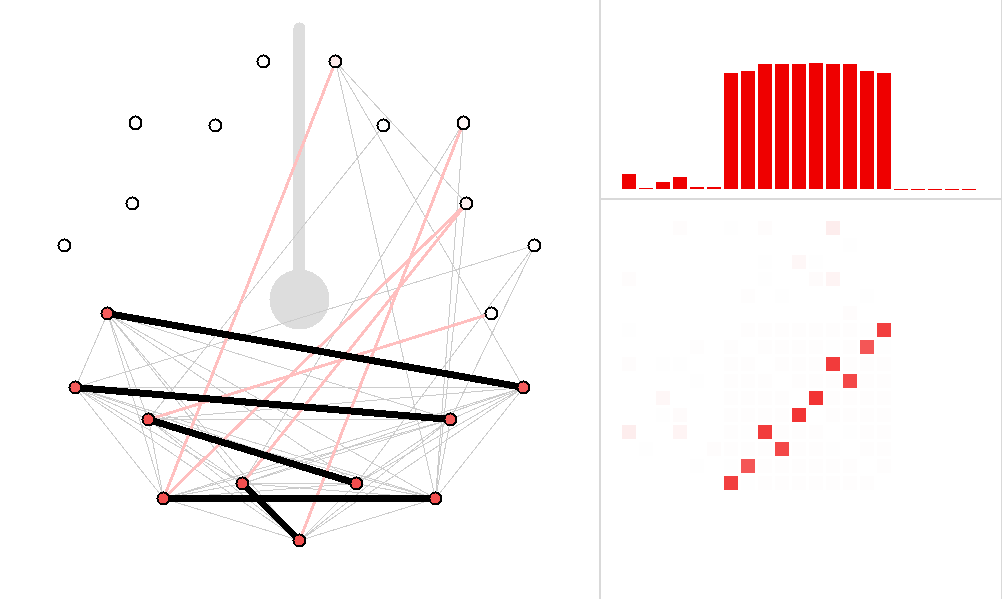} & $\;\;\;$ &\includegraphics[width=8cm]{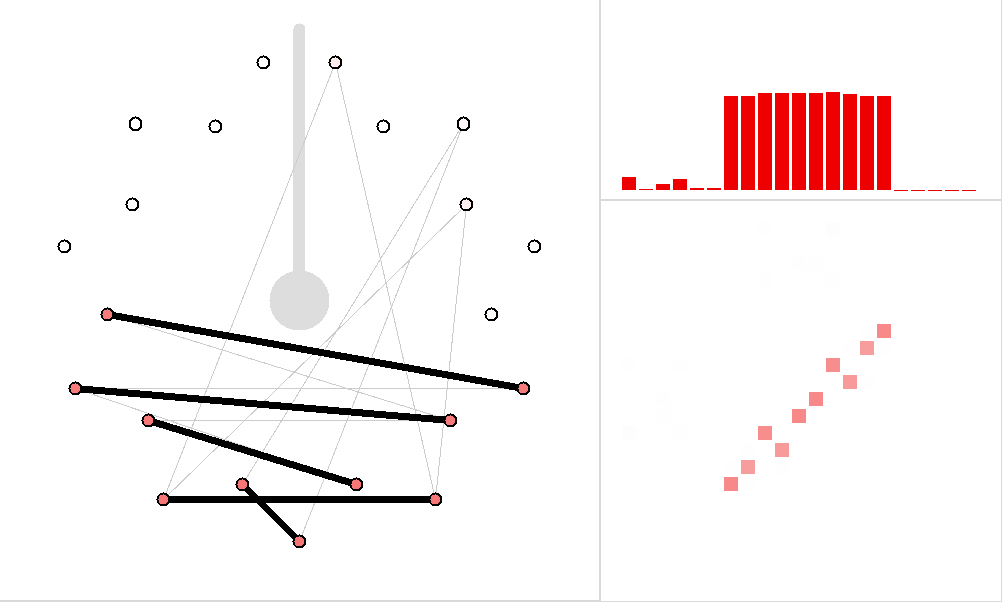}\\
\end{tabular}
\end{center}
\end{figure}

Fig.~\ref{S1analysis} shows the results for the first excited ($S=1$) state in its two $M_s=1,0$ components.
The overall situation is similar to the ground state,  except that here one pair of orbitals remains open shell ``ferromagnetically'' triplet-coupled. 
The spin-including mutual informations differ for the two $M_s$ components,
while the spin-free ones are identical and a similar ``weakening'' of the  mutual information and orbital entropies as in the ground state happens.
The situation is analogous for the other excited states, with the most dramatic spin-free ``simplification'' occurring 
for the $M_s<S$ components of the ``totally ferromagnetic'' $S=5$ excited state (analogous to the $S=5/2$ state of the single-iron complex, cf. Fig.~\ref{fecomplexanalysis}).
Rather than showing figures for all the cases, we summarize the results in Tab.~\ref{fe2s2complextable}.
The somewhat less than perfect $M_s$-invariance of the spin-free quantities for higher DMRG roots
corresponds to the truncation of the bond dimension, which was limited to 2048 for all states and is more limiting 
to the low-$M_s$ states being higher roots. 
In some cases we did not achieve convergence at all for the higher roots.
Nevertheless, the total spin-free entropy $\tilde{S}_{\rm tot}$ shows the same increasing trend towards low spin states as in the complex with single iron atom.
On the other hand, the total spin-free mutual information  $\tilde{I}_{\rm tot}$ is very similar (around 2.2) for
states with $S=0-4$ and decreases to 0.13 only for the highest spin state $S=5$. This is in line with the observation
from Figs.~\ref{S0analysis},\ref{S1analysis}, which show strong pairwise  \Blue{correlation} between orbitals from the two iron atoms,
both for the ground state $(S=0)$ and first excited state $(S=1)$. This  \Blue{correlation} ``survives'' in the spin-free analysis
and is thus not caused by spin coupling, but reflects the inherent strong correlation in the system.

\begin{table}
\caption{
\label{fe2s2complextable}
Spin-free and spin-including total \Blue{quantum informations} $\tilde{S}_{\rm tot}$, $S_{\rm tot}$ and mutual informations  $\tilde{I}_{\rm tot}$, $I_{\rm tot}$ for three states of the  [Fe$_2$S$_2$(SCH$_3$)$_4$]$^{2-}$  complex and their $M_s$ components.
}
\begin{tabular}{ccrrrr}
\hline
$S$ & $M_s$ &  $\tilde{S}_{\rm tot}$ & $S_{\rm tot}$ &  $\tilde{I}_{\rm tot}$ & $I_{\rm tot}$ \\
\hline
0 & 0 & 11.028 & 13.642 & 4.456 & 8.470\\
1 & 1 & 10.925 & 13.346 & 4.342 & 8.276 \\
1 & 0 & 10.920 & 13.854 & 4.340 & 8.224 \\
2 & 2 & 10.490 & 12.210 & 4.374 & 7.854 \\
2 & 1 & 10.488 & 13.612 & 4.374 & 7.560 \\
2 & 0 & 10.473 & 14.021 & 4.362 & 7.948 \\
3 & 3 &  9.303 & 10.134 & 4.398 & 6.910 \\
3 & 2 &  9.302 & 12.367 & 4.400 & 8.400 \\
3 & 1 &  9.300 & 13.402 & 4.398 & 10.158 \\
3 & 0 & \multicolumn{4}{c}{not conv.}\\
4 & 4 &  6.717 &  6.791 & 4.404 & 5.504 \\
4 & 3 &  6.717 &  9.761 & 4.404 & 10.578 \\
4 & 2 &  6.716 & 11.239 & 4.404 & 13.840 \\
4 & 1 &  6.618 & 11.966 & 4.262 & 15.914 \\
4 & 0 &  \multicolumn{4}{c}{not conv.}\\
5 & 5 &  0.684 &  0.687 & 0.262 & 0.642 \\
5 & 4 &  0.684 &  3.936 & 0.262 & 7.370 \\
5 & 3 &  0.684 &  5.689 & 0.260 & 12.030 \\
5 & 2 &  0.684 &  6.793 & 0.260 & 15.472 \\
5 & 1 &  \multicolumn{4}{c}{not conv.}\\
5 & 0 &  \multicolumn{4}{c}{not conv.}\\
\hline
\end{tabular}
\end{table}

\section{Conclusions}

We have defined spin-free analogues of orbital entropy, pair entropy, and mutual information and investigated their properties.
We have shown that the spin-free counterparts cannot be bigger than the original spin-including quantities. 
Typically they are smaller, which indicates that a part of the strong correlation is actually a ``static'' correlation due to spin couplings.
We have shown and numerically verified the invariance of the spin-free entropy and mutual information with respect to the $M_s$ component of the
investigated state and have observed that the spin-free quantities are close to the original ones for the maximum $M_s$ component of the high-spin case.
We studied two realistic examples of complexes with iron-sulfur bonds which exhibit strong correlation and by comparison of the spin-free
and spin-including correlation measures we were able to identify to which extent the correlation is due to spin couplings.
A trend of increase of the total spin-free entropy  with decreasing spin of the state has been observed.
Since the computation of the spin-free  \Blue{correlation} analysis does not require any significant additional computational cost, 
we think that it can be routinely used as a useful supplement to the analysis using the ``traditional'' spin-dependent orbital entropy and mutual information.

\section*{Acknowledgments}

This paper is dedicated to Professor Piotr Piecuch on the occasion  of his 65th birthday.

The work has been supported by the Advanced Multiscale Materials for Key Enabling Technologies project of the Ministry of Education, Youth, and Sports of the Czech Republic. Project No. CZ.02.01.01/00/22\_008/0004558, Co-funded by the European Union.
The computational time was supported by the Ministry of Education, Youth and Sports of the Czech Republic through the e-INFRA CZ project (ID:90254).
The author thanks Dr. Libor Veis for adding ASCII printout of the reduced density matrices to the MOLMPS DMRG code.

\bibliographystyle{achemso}
\bibliography{entropy,ors,cc,perspective}

\end{document}